\documentclass[iop]{emulateapj}

\usepackage{graphicx}
\usepackage{txfonts}
\usepackage{hyperref}
\usepackage{url}
\usepackage{natbib}
\usepackage{amssymb}
\usepackage{xcolor}

\bibpunct{(}{)}{;}{a}{}{,}

\newcommand{\ap}{\ensuremath{\approx}}

\newcommand{\pow}[1]{\ensuremath{\times10^{#1}}}

\begin{document}

\title{A resolved and asymmetric ring of PAHs\\within the young circumstellar disk of IRS 48}

\shortauthors{Schworer et al.}
\author{
  Guillaume~Schworer\altaffilmark{1,2}, 
  Sylvestre~Lacour\altaffilmark{1,3}, 
  Nuria~Hu\'elamo\altaffilmark{4}, 
  Christophe~Pinte\altaffilmark{5,6} 
  \\
  Ga\"el~Chauvin\altaffilmark{5}, 
  Vincent~Coud\'e~du~Foresto\altaffilmark{1},
  David~Ehrenreich\altaffilmark{7},
  Julien~Girard\altaffilmark{8},
  Peter~Tuthill\altaffilmark{2}
}


  \affil{\altaffilmark{1}LESIA, Observatoire de Paris, PSL Research University, CNRS, Sorbonne Universités, \\UPMC Univ. Paris 06, Univ. Paris Diderot, Sorbonne Paris Cit\'{e}, France}
  \affil{\altaffilmark{2}Sydney Institute for Astronomy, School of Physics, The University of Sydney, NSW 2006, Australia}
  \affil{\altaffilmark{3}Cavendish Laboratory, University of Cambridge, JJ Thomson Avenue, Cambridge CB3 0HE, UK}
  \affil{\altaffilmark{4}Dpto. Astrof\'{\i}sica, Centro de Astrobiolog\'{\i}a (INTA-CSIC), ESAC Campus, PO Box 78, 28691, Villanueva de la Ca\~nada, Spain}
  \affil{\altaffilmark{5}Univ. Grenoble Alpes, IPAG, F-38000 Grenoble, France\\CNRS, IPAG, F-38000 Grenoble, France}
  \affil{\altaffilmark{6}UMI-FCA, CNRS/INSU France (UMI 3386), and Departamento de Astronom{\'\i}a, \\Universidad de Chile, Casilla 36-D Santiago, Chile}
  \affil{\altaffilmark{7}Observatoire de l'Universit\'e de Gen\`eve, 51 chemin des Maillettes, 1290 Versoix, Switzerland}
  \affil{\altaffilmark{8}European Southern Observatory, Alonso de Cordova 3107, Casilla 19001 Vitacura, Santiago 19, Chile}

\begin{abstract}
For one decade, the spectral-type and age of the $\rho$ Oph object IRS-48 were subject to debates and mysteries. Modelling its disk with mid-infrared to millimeter observations led to various explanations to account for the complex intricacy of dust-holes and gas-depleted regions. We present multi-epoch high-angular-resolution interferometric near-infrared data of spatially-resolved emissions in its first 15AU, known to have very strong Polycyclic Aromatic Hydrocarbon (PAH) emissions within this dust-depleted region. We make use of new Sparse-Aperture-Masking data to instruct a revised radiative-transfer model where SED fluxes and interferometry are jointly fitted. Neutral and ionized PAH, Very Small Grains (VSG) and classical silicates are incorporated into the model; new stellar parameters and extinction laws are explored. A bright (42L$_{\odot}$) central-star with A$_v$=12.5mag and R$_v$=6.5 requires less near-infrared excess: the inner-most disk at \ap1AU is incompatible with the data. The revised 
stellar parameters place this system on a 4 Myr evolutionary track, 4 times younger than previous estimations, in better agreement with the surrounding $\rho$ Oph region and disk-lifetimes observations. The disk-structure converges to a classical-grains outer-disk from 55AU combined with a fully resolved VSG\&PAH-ring, at 11-26 AU.
We find two over-luminosities in the PAH-ring at color-temperatures consistent with the radiative transfer simulations; one follows a Keplerian circular orbit at 14AU. We show a depletion of a factor \ap5 of classical dust grains compared to VSG\&PAH: the IRS-48 disk is nearly void of dust-grains in the first 55 AU. A 3.5M$_{Jup}$ planet on a 40AU orbit qualitatively explains the new disk-structure.
\end{abstract}


\maketitle


\section{Introduction}

The dust- and gas-rich disks surrounding numerous pre-main-sequence (PMS) stars are of key interest for unveiling how planetary system are formed; they are the initial conditions for planetary formation. Protoplanetary disks have a rich structure, with different physics playing a role in different regions of the disk. The dynamic ranges involved span two to five orders of magnitudes on spatial scales, orbital times, temperatures, and much more in dust or gas densities. The extreme dynamic ranges involved in the structure and composition of these objects mean that very different observational techniques have to be combined together to probe their various regions.

While spatially resolved observations give a direct localization of emitting regions, allowing precise disambiguation of structures in the spatial domain, a somewhat similar approach exists in the spectral domain through the analysis of the relative brightness distribution of parts (or all of) the disk. The discovery of a medium-age sub-class of young circumstellar disks, known as transition disks, characterized by a distinctive dip in their infra-red Spectral Energy Distribution (SED) suggests that a partially evacuated gap exists in the inner region of the protoplanetary disk~\citep[e.g. ][]{Calvet02}. The profound implications for studies of planetary formation have become increasingly apparent with the confirmation of the disk-gap architecture by sub-millimeter measurements and optical interferometry~\citep[e.g. ][]{Andrews11}.

\begin{table}[!]
\caption{Main literature data of $\rho$ Oph IRS-48}
\begin{center}
\begin{tabular}{lll}
\hline\hline
Parameter &  Value & Ref. \\
\hline
Stellar position    & 16$^{\rm h}$27$^{\rm m}$37$.\!\!^{\rm s}$18  &  B14 \\
(J2000)             & -24$^{\circ}$ 30\arcmin 35.3\arcsec  & \\
Distance            & 120 pc & L08 \\
Inclination         & i=50$^{\circ}$ & G07, B14 \\
Systemic velocity   & 4.55 km.s$^{-1}$ & vdM13 \\
Position angle      & PA=96$^{\circ}$ & G07, B14 \\
Stellar type        & A0$^{+4}_{-1}$ & B12 \\
Stellar Temp.       & 9000$\pm$550 K & B12 \\
\hline
\end{tabular}\par
\end{center}

B14$=$\citet{Bruderer14}, L08$=$\citet{Loinard08}, B12$=$\citet{Brown12a}, vdM13$=$\citet{vanderMarel13}
\label{table:main_data}
\end{table}

IRS-48 is a spectacular candidate where the delicate balance of such phenomena is currently questioned. Also referred as $2MASS~J16273718-2430350$, $GY~304$ or $WLY~2-48$, it is an highly extinct A0$_{-1}^{+4}$ star, part of the $\rho$ Oph cloud L1688; Table~\ref{table:main_data} sums up its main data.

First listed in~\citet{Wilking89}, it shows a substantial far-infrared and millimeter excess, pointing towards a classification as transition disk. Since the first resolved images obtained by~\citet{Geers07} in the near- and mid-infrared, and the first substantial evidence of Polycyclic Aromatic Hydrocarbon (PAH) emission lines in~\citet{Geers07b}, IRS-48 has been an object of interest and mysteries. While 8.6 through 11.3 $\mu$m images showed high and unresolved PAH fluxes, the 18.7 $\mu$m image uncovered a purely thermal and asymmetric outer-disk from a radius of 55 AU. Using the VLT-CRIRES spectrograph to observe the 4.7 $\mu$m CO fundamental rovibrational band,~\citet{Brown12a} imaged a 30 AU thin ring. Additionally, they solve the spectral-type puzzle of the central star and set a luminosity of 14.3 L$_{\odot}$ combined with an extinction of A$_v$=11.5, positioning the object on a -- suspiciously old -- 15 Myr evolutionary track.

Subsequent ALMA and VLA observation at 0.44, 1.3, and 8.8 mm carried by~\citet{vanderMarel13,vanderMarel16} unveiled a millimeter-grains asymmetry at a radius of 63 AU, in the southern extents of the disk. They postulate a planet of 10 M$_{Jup}$ located at \ap18 AU triggering a vortex-shaped dust-trap. Further work on PAH by~\citet{Maaskant14} using unresolved spectral energy distribution data, revealed that the near- and mid-infrared spectrum of IRS-48 is dominated by a mixture of neutral and ionized PAH, that they postulate to arise from an extended region between the inner-most disk at 1 AU to the outer-disk at 55 AU.

Most recent studies by~\citet{Bruderer14} showed that a single gas-depletion in the inner disk could not explain the flux profiles of $^{12}$CO and C$^{17}$O lines measured by ALMA with a resolution of 30 AU. They proposed a slight gas depletion at \ap20-50 AU in addition to a complete depletion inside 20 AU.~\citet{Follette15} report the first reflected light H- and Ks-band direct images of the outer 55 AU disk, and question the abundance of the innermost-disk at \ap1 AU using SED modelling, although their fit suffers from over-luminous silicate features in the 9-18 $\mu$m window.

A number of mechanisms have been proposed to cause gaps in proto-planetary disks, including extensive grain growth~\citep{Dullemond05}, photo-evaporation \citep{Clarke01}, binarity \citep{Ireland08}, and tidal barrier created by dynamical interaction with low-mass disk objects~\citep[e.g. ][]{Bryden99}. These different mechanisms can be distinguished by studying the distribution of the gas and dust within the gaps: a large (stellar) companion or photo-evaporation would almost completely evacuate the inner regions while a less massive planetary companion would allow gas and small dust grains to exist within its orbit \citep{Lubow99}. Furthermore, in the latter case, the measurement of the size and distribution of this material would allow the orbit and mass of the planetary companion to be constrained.

Key grain-coagulatation theories~\citep[][ for a review]{Williams11} are able to explain growth from submicrons to millimeters, and then hectometer to planets. Several possible solutions have been proposed to account for the formation of bodies up to meter-sizes, such as turbulent vortices \citep{Heng10}. Another mystery regarding grain-coagulation exists for the smallest grains, that can arguably be considered as large molecules: PAH, and in a lesser extent, Very Small Grains (VSG). It has been argued that PAH either may take part in the dust coagulation process~\citep{Dullemond07} or are replenished by mixing processes in the disk~\citep{Siebenmorgen10}. However, observational evidence of the role of PAH in these processes is not conclusive and it is presently unclear how important these routes are.
The question remains whether these very small particles take part in grain-coating, in the main grain-growth process (and their presence in many disks are to be explained by some replenishment processes), or if their intrinsic properties prevents them to grow (and their absence in many disks are to be explained by some other assimilation processes).
\citet{Gorti08} find that the abundance of PAH and small grains in disks is critical to the temperature profile. Indeed, they show that if some PAH are present, their contribution to disk heating via grain photoelectric emission can be larger by a factor of \ap2 than X\-ray heating at r\ap10 AU.

This paper reports the first direct detection of the full extents of a PAH ring in a young circumstellar disk, presents a revised model for the IRS-48 object to explain the rich and complex dust- and gas-environment observed from near-infrared to centimeter wavelengths, and sets limits on how much silicates grains -- hence replenishment -- is to be expected in the PAH and VSG ring.

\section{Observations and Data Reduction}

\subsection{Observation Strategy}
In this work, we make use of new data sets acquired using the instrument NaCo (abbreviation for Nasmyth Adaptive Optics System (NAOS) \& Near-infrared Imager and Spectrograph (CONICA)) commissioned on the Very Large Telescope (VLT) at the Paranal Observatory of the European Southern Observatory (ESO). This data was acquired in different near-infrared bands and at four epochs over a two-years period; these sets are detailed in Table~\ref{table:observation_nacosam}.
At that time, NaCo was integrated at the Nasmyth platform of the UT4 8.2m-telescope. We used this instrument in Sparse-Aperture Masking (SAM) and full-pupil imaging modes. For both modes, given the large reddening of the target ($>$16mag in the visible), we use the NaCo built-in infrared wavefront sensor for the Adaptive Optics (AO) correction.

\begin{table}[!]
\caption{Observations made with NaCo}
\begin{center}
\begin{tabular}{ l c c c }
  \hline\hline
  Date & Band & Calibrators & Seq. \\ \hline
  14 Mar. 2011 & Ks$^{(1,2)}$ & Elia 2-35 \& 2-37 & 4 \\ 
  14 Mar. 2011 & Lp$^{(1,2,3)}$ & Elia 2-37$^{(1,2)}$ & 4 \\ 
   & & Hip 86311$^{(3)}$ & \\
  14 Mar. 2011 & Mp$^{(1,2)}$ & Elia 2-15 & 2 \\ 
  1 Sept. 2011 & Lp$^{(1)}$ & Elia 2-37 & 2 \\ 
  6 Mar. 2012 & Lp$^{(1)}$ & Elia 2-37 & 3 \\ 
  25 Mar. 2013 & Lp$^{(1)}$ & Elia 2-37 \& 2-11 & 4 \\ 
\end{tabular}\par
\end{center}
$^{(1)}$: SAM\\$^{(2)}$: Full-pupil image\\$^{(3)}$: Spectra\\A calibration sequence (``Seq.'' in the headline, also called ``bracket'') is a unitary interferometric measurement following the observation pattern ``Calibrator--Science Target--Calibrator''\\Single-frame exposition time is 0.4 sec for both SAM and imaging modes.
\label{table:observation_nacosam}
\end{table}

\subsection{Direct Imaging and Spectroscopy}

\begin{figure*}[!]
\centering
  \includegraphics[width=0.5\hsize]{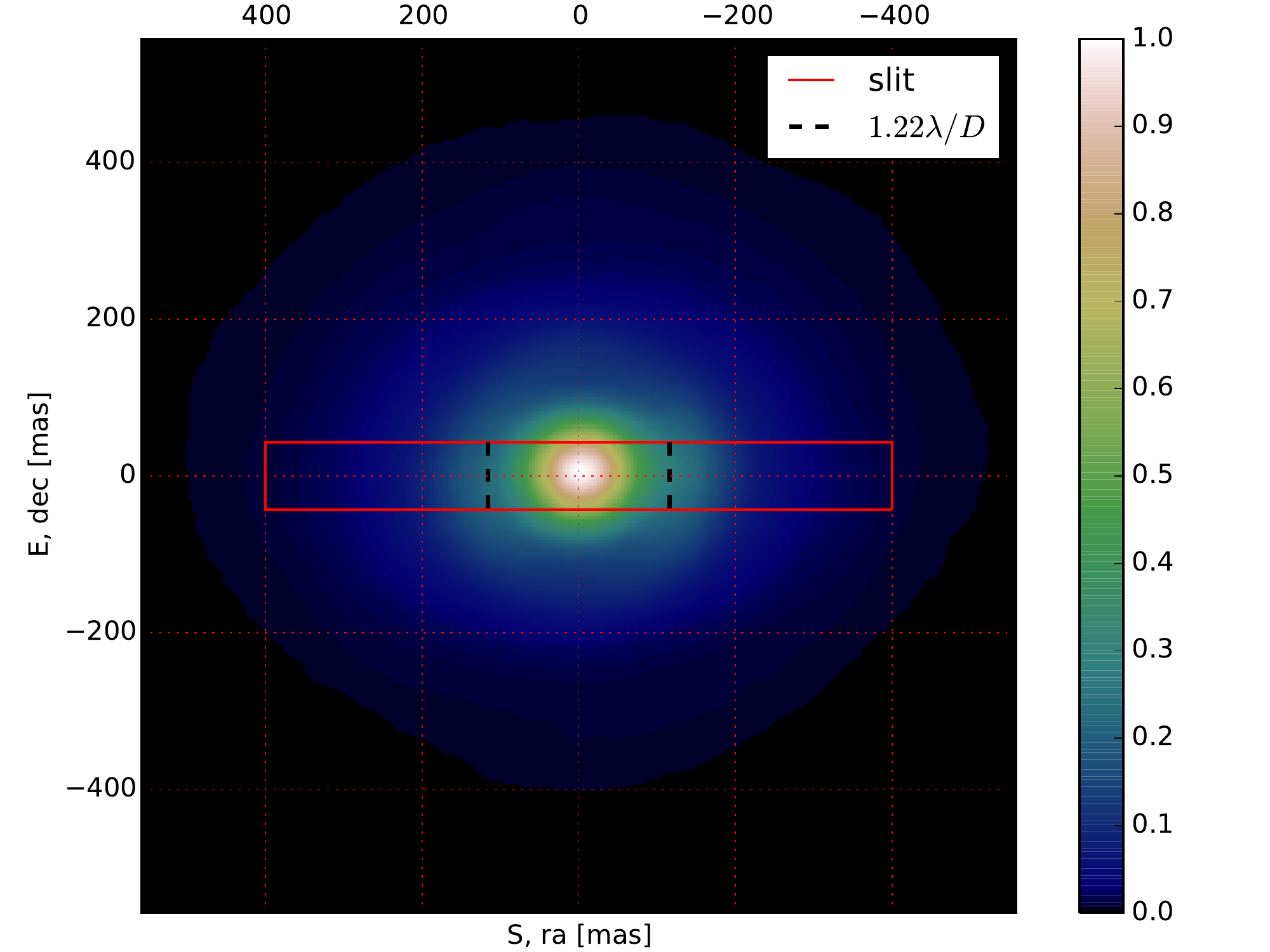}%
  \hfill%
  \includegraphics[width=0.5\hsize]{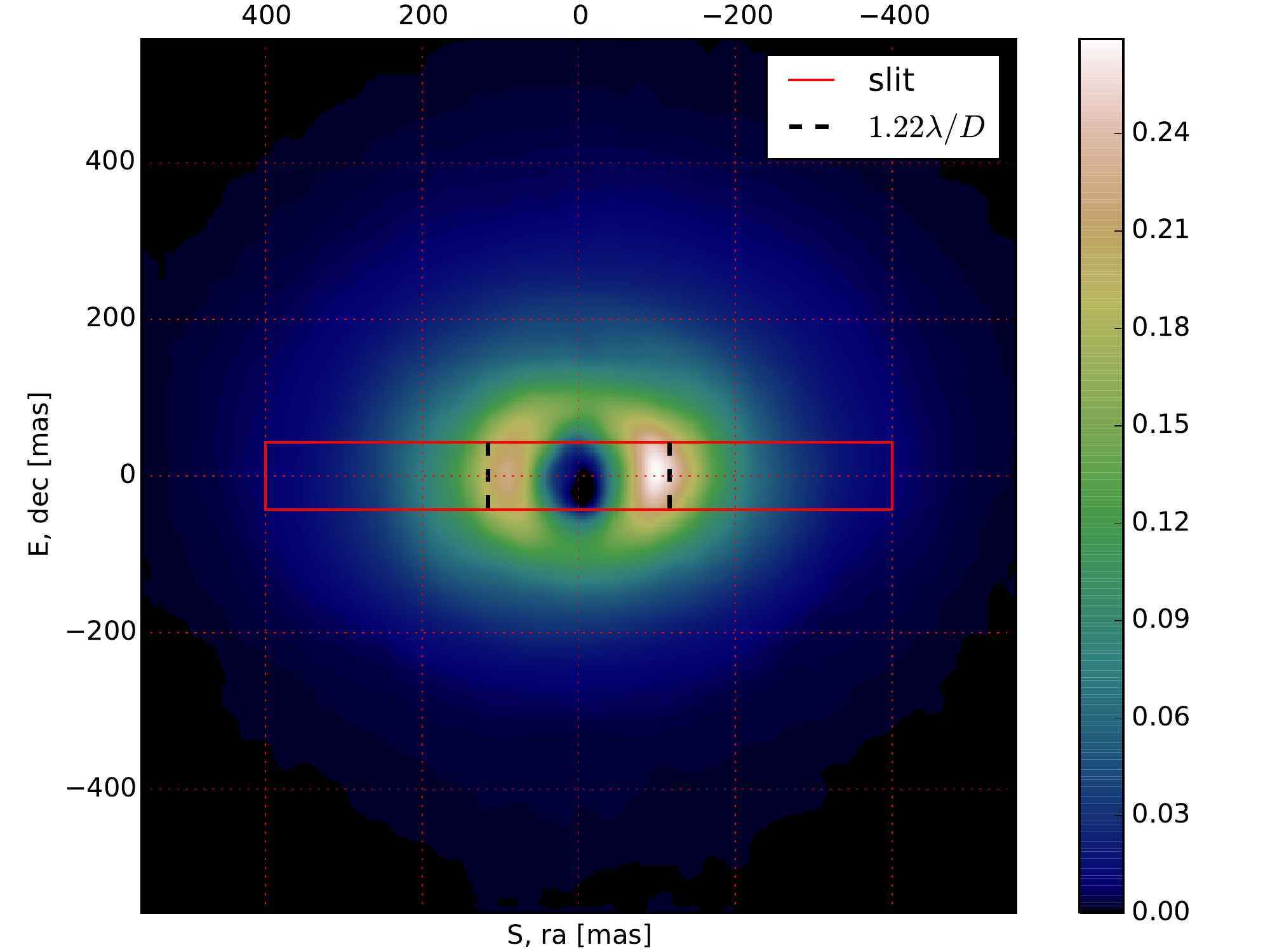}\par
  \includegraphics[width=\hsize]{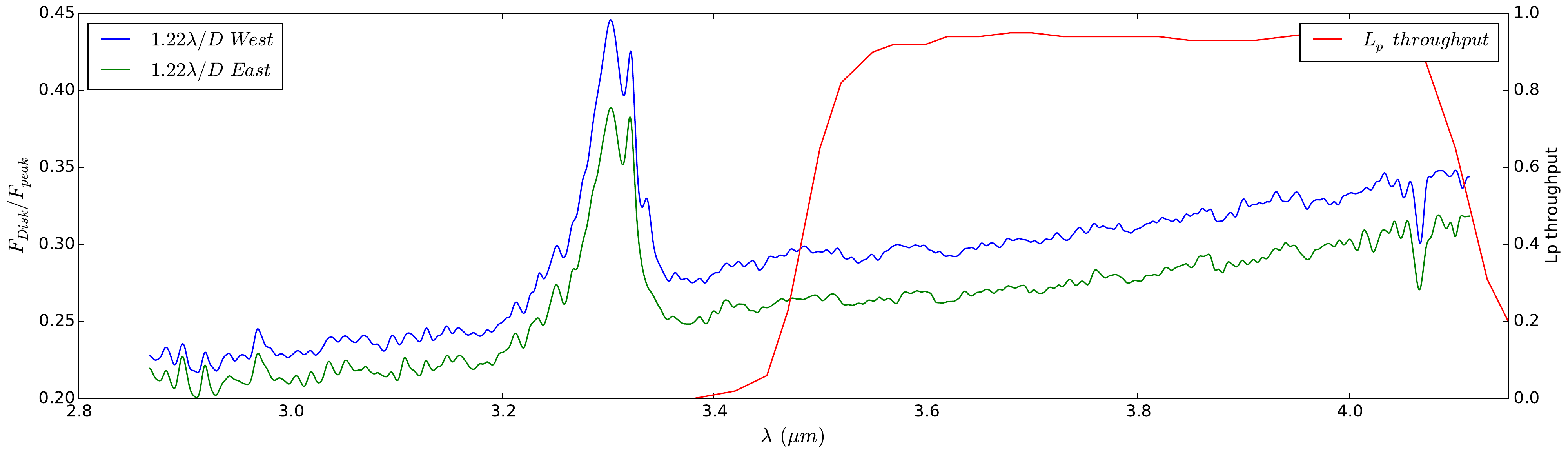}
\caption{Full-pupil image in Lp-band (top left), and PSF-subtracted (top right) in normalized flux (peak of the un-subtracted image). The 86 mas slit height is overlaid in red, while the black dashed-lines lying at \ap115 mas illustrates the first null of the central star, at 1.22$\lambda$/D. The displayed spectra show the disk-spectra taken at the East and West stellar-nulls, divided by the spectra taken at the peak intensity. Prominent PAH emission is seen from the disk at 3.3 $\mu$m, and $Br\alpha$ from the star at 4.05 $\mu$m. The same East-West asymmetry is seen as the one observed on the image.}
\label{figure:img}
\end{figure*}

At epoch 1, full-pupil images were acquired in Ks-, Lp- and Mp-bands, as well as spectra between 2.87 and 4.11 $\mu$m, with $R_{average}\ap800$. The Lp-band image is displayed in Figure~\ref{figure:img} both with and without PSF-subtraction. One can easily see the East-West elongated nature of the circumstellar emission, at \ap100 milli-arcsecond (mas), as well as an asymmetry in the disk semi-major axis; the Western extension is \ap1.15 times brighter than the Eastern one. However, the \ap100 mas spatial scale is similar to the resolution criterion $\lambda$/D=96 mas at these wavelengths, which enormously complexifies the use of these images.

The slit of the spectrograph was orientated along the East-West direction as displayed in the latter Figure. The pixel-scale of the detector is constant over the whole spectra: 1.6 \AA~$\times$2.7 mas. The slit height is 86 mas; given the short scale extents of the disk and the high inclination (\ap50$^{\circ}$), we expect a large contamination of the disk at the stellar position (assumed to be the located at the peak intensity in the spectra). Similarly, given the full-width-half-maximum of the PSF at \ap50 mas, we expect a large contamination of the star in the disk. To minimize the stellar contamination on the disk, we make use of the spectra at the first null of the star, 1.22$\lambda$/D\ap115 mas. The PSF-reference image shows a very low stellar-flux in the null, \ap3\% of the peak-flux; this highlights the high strehl ratio of the observation.

The spectra in Figure~\ref{figure:img} shows the flux of the disk -- taken at the first stellar null -- relative to the peak flux of the spectra. The extraction aperture is three pixels for the peak, and one pixel interpolated at 1.22$\lambda$/D for the null. One can see very strong PAH emission features at 3.3 $\mu$m, characteristic of neutral PAH: in this emission line the disk-to-peak contrast jumps by nearly a factor of two, compared to the disk-to-peak continuum contrast. Note that this emission line is not recorded in the interferometric data or full-pupil images given that it is outside the Lp-band filter bandpass, see Figure~\ref{figure:img}. Although both East and West spectra are very similar in shape, the Western disk-extension is constantly -- at all spectrum wavelengths -- 1.15 times brighter than its Eastern region. The same West-East asymmetry was measured on the full-pupil image.

\subsection{Near-Infrared Sparse-Aperture Masking}

The most interesting data sets are SAM data in that they offer an angular resolution down to a few tens of mas, equivalent to a few AU given the distance of 120 parsec, as shown in Table~\ref{table:res_fov}. Except for the longest baselines of ALMA ($\gtrsim$1 km), such resolution cannot be reached with any other instrument nor any other NaCo mode (coronographic, full-pupil, etc). Moreover, new extreme adaptive-optics (AO) systems such as SPHERE usually have their wavefront sensors in the visible and cannot close the AO loop on this extremely extinct target ($>$16mag in the visible), which means that NaCo-SAM is the only instrument able to observe IRS-48 at such resolution in visible or in the infrared (IR).

SAM, also called Non-Redundant Aperture Masking (NRM) is an observational technique in which a single mirror is transformed into a multi-pupil Fizeau interferometer. A mask placed in the pupil plane of the telescope transforms the usual Airy point-spread function (PSF) into a pattern made of multiple fringes. The technique consists in measuring the amplitude and phase of these interferometric fringes so that information can be retrieved down to a fraction of a resolution element of the longest baseline of the mask (typically 0.5$\lambda/B_{max}$\ap50 mas at 4 $\mu$m). However, the interferometric field of view remains very limited, in practice of order 2.5$\lambda/B_{min}$\ap250 mas at 4 $\mu$m (although a somewhat lower sensitivity exist until \ap6$\lambda/B_{min}$). Using closure phases (T3) -- an estimator robust to atmospheric turbulence and optical aberrations \citep{Jennison58,Monnier00} -- allows high-contrast capability without the inevitable ``blind-spot'' of other techniques such as coronography.

The observation strategy for SAM mode was designed to intertwine the science target IRS-48 between two blocks on an interferometric calibrator, here Elia 2-35, 2-37 or 2-15, as detailed in Table~\ref{table:observation_nacosam}. The mask used was a 7-hole non-redundant mask with baselines ranging from 1.78 to 6.43 meters, allowing the simultaneous measurement of 21 visibility-squared (VIS2) and 35 closure phases (T3, among which 15 are independent). In full-pupil imaging mode, we used the same calibrators to have PSF references.

The data was processed using the $SAMP$ pipeline \citep{Lacour11}. It includes sky subtraction, bad pixel subtraction and fringe fitting. T3 and VIS2 measurements are calibrated by the two calibrators for each bracket.

\begin{table}[!]
\caption{NaCo filters characteristics, interferometric 0.5 $\lambda/D$ criteria and field of view with respect to filters at central wavelengths, given a target distance of 120 parsec.}
\centering
\begin{tabular}{ l c c c }
  \hline\hline
   & Ks & Lp & Mp \\ \hline
  Central wavelength ($\mu$m) & 2.18 & 3.8 & 4.78 \\
  Bandwidth ($\mu$m) & 0.35 & 0.62 & 0.59 \\ \hline
  \multicolumn{4}{c}{Smallest Baseline, 1.78m} \\
  $0.5~\lambda/D$ (AU) & 15.2 & 26.6 & 33.5 \\
  Field of view (AU) & 190 & 326 & 544 \\ \hline
  \multicolumn{4}{c}{Longest Baseline, 6.43m} \\
  $0.5~\lambda/D$ (AU) & 4.2 & 7.4 & 9.2 \\
  Field of view (AU) & 52.6 & 90.8 & 150 \\ \hline
\end{tabular}
\label{table:res_fov}
\end{table}

Calibration to PSF-reference help reduce systematic errors in the data, but hardly corrects all of them. A fitting mechanism to account for that is specifically developed, refer to Section~\ref{section:t3morph}. The error-bars on the data are obtained from statistical measurements on the data, which account for random errors. T3 errors they average \ap0.5$^{\circ}$ in Lp-band and \ap1.7$^{\circ}$ in Ks- and M-band.

\subsection{Spectral Energy Distribution}

In addition to the multi-wavelength interferometric data listed above, we compiled SED fluxes for IRS-48, from the literature. All data are listed in Table~\ref{table:literature_photometry} with the corresponding references. As can be seen, the photometry from literature has a broad coverage from visible-blue to mm wavelengths. No measurements in the ultraviolet exist because of the very high extinction of the source; measurements in cm-wavelengths are discarded given that they trace cold and large-scale structures outside the region of sensitivity and interest of our observations. For the fitting exercise, the errors on photometry fluxes were assumed to be of 5\% if not specified in the literature.

The correction of the interstellar extinction was made using~\citet{Cardelli89}, improved by~\citet{oDonnell94} work in the visible. These extinction curves are parametrized with R$_v$ (reddening slope) and A$_v$ (absorption in V band) such that A$_v\equiv$ R$_v$*Extinction(B$_{mag}$-V$_{mag}$). The initial values are taken from \citet{Brown12a}, A$_v$=11.5 and R$_v$=5.5.

\begin{table}[!]
\caption{Previously published photometry and spectra of $\rho$ Oph IRS-48.}
\begin{center}
\begin{tabular}{ c c c }
  \hline\hline
  Wavelength(s) & Instrument & Reference \\
  ($\mu$m) & & \\
  \hline
  0.43, 0.64 & NOMAD & Z05 \\
  0.65, 0.8 & Hydra & E11 \\
  1.2, 1.6, 2.2 & 2MASS & C03 \\
  3.4, 4.6 & WISE & W10 \\
  3.6, 4.5 & Spitzer IRAC & vK09 \\
  5.9-36.9 & Spitzer IRS & MC10 \\
  60-181 & Herschel PACS & F13 \\
  450 & ALMA & vdM13 \\
  850 & SCUBA & A07 \\
  880, 1300 & SMA & B12b \\
  \hline
  \end{tabular}\par
\end{center}
Z05$=$\citet{Zacharias05}, E11$=$\citet{Erickson11}, C03$=$\citet{Cutri03}, W10$=$\citet{Wright10}, vK09$=$\citet{vanKempen09}, MC10$=$\citet{McClure10}, F13$=$\citet{Fedele13}, A07$=$\citet{Andrews07}, B12b$=$\citet{Brown12b}, vdM13$=$\citet{vanderMarel13}
\label{table:literature_photometry}
\end{table}

\section{Radiative Transfer Model}

\subsection{\textit{Old} Model}

We take over the disk structure presented in~\citet{Bruderer14}. Given the prominence of PAH in the SED of IRS-48, we merge into this disk-model the results of~\citet{Maaskant14} on ionized PAH in several young disks. We obtain a disk-structure in three parts: 1) an inner-most dusty disk between 0.4 and 1 AU, 2) an outer dusty disk between 55 and 160 AU, and 3) a PAH disk in-between. For the sake of simplicity, we call this disk model the \textit{old} model. Its main parameters can be seen in Table~\ref{table:mcfost_model}. This model represents the literature knowledge of IRS-48.

We use MCFOST, a parallelized 3D radiative transfer code based on a Monte-Carlo method~\citep{Pinte06}, able to output SED data and monochromatic images when given an input parameter file. After a few minor adjustments, the fit to the SED literature data is remarkable. However, this model or any small variation of it fails to explain the VIS2 data in either Lp- or Mp-band (see Figures~\ref{figure:VIS2_bl}). This highlights the degeneracy of SED-fitting exercises.

Indeed, the inner dust-disk between 0.4 and 1 AU emits very strongly in NIR, while the PAH grains further out are much dimmer. This creates a very high unresolved-flux (inner-disk and star) compared to the resolved-flux from the PAH in the \ap9-55 AU region. Hence, the inner regions of the \textit{old} model disk appears spatially too small, mostly unresolved and the resulting visibilities are close to unity: much larger than the VIS2 data at all baselines. Given the difference of the VIS2 values, the resolved-flux depletion is estimated to \ap2.7 for baselines $\gtrsim$3 meter.

The only solution to make the old model fit the VIS2 is to incorporate a very high flux in the \ap9-35 AU region -- and in this region only --, to make up for the contrast ratio of \ap2 in Lp-band between unresolved and resolved flux.

This implies changing fundamentally the structure, hence the nature, of the disk around IRS-48.

\subsection{New model: Fitting Strategy}

In our data, VIS2 measurements are not excessively sensitive to asymmetries, essentially because of their relatively large error-bars. Hence, they encode a point-symmetrical and smooth disk-structure and inform on high-level morphology. We make use of the VIS2 data in Lp-band from all epochs and the Mp-band taken at epoch 1, as well as the SED fluxes from the literature (refer to Table~\ref{table:literature_photometry}). Ks-band from epoch 1 are discarded given that their large uncertainties would not set meaningful constraints on the disk structure. The Lp-band VIS2 data does not show time-variation along the four epochs.

We decided to perform a manual initial exploration of the parameters to identify the most impacting ones. We refine the fit later on with both a genetic algorithm and low-dimensional grids. Genetic algorithm allows for a very fast convergence for cases with a large number of dimensions where basic MCMC would fail. Grid calculations only come in a second time and ensure a full exploration of a small subset of the entire parameter space. Verifying smoothness for the outputs of the modelling over these small grid-subsets allowed us to interpolates the results and thus minimize the processing time for these grids. The final convergence was achieved by eye.

For the purpose of measuring visibilities on these MCFOST images, we choose their pixel resolution in ``mas per pixel'' to display the full interferometric field of view in a constant width of 1001 pixels. Also, in order to account for bandwidth smearing, we generate 15 monochromatic images linearly spaced in wavelength over the bandwidth of each filter (see Table~\ref{table:res_fov}) and combine them into a unique polychromatic image.

We perform a SED-fitting ``inside-out'', or ``in increasing wavelengths'', so that constraints and parameters obtained from fitting interior structures can remain (mostly) unchanged when fitting structures located further away from the star.

\subsection{New model: Initial Configuration and Assumptions}
We assume an axial-symmetry perpendicular to the plane of the disk as well as plane symmetry above and below the plane of the disk. We adopt the standard flared disk prescription \citep{Shakura73}, described in cylindrical coordinates ($r$, $z$) such that the density profile is given by:
\begin{equation}\label{eqn:dustdensity}
    \sum(r,z) = \sum_0~r^p \exp\Bigg(-\frac{1}{2}\bigg(\frac{z}{h}\bigg)^2\Bigg),
\end{equation}
where $\sum_0$ is a density-normalization constant, $p$ is the surface density exponent, and $h$ is the disk scale height; $h$ varies with radius as $h = h_0(r/r_0)^\beta$ with $\beta$ the flaring exponent ($\beta>0$, of order \ap1) and $h_0$ the scale height at radius $r_0$.

The SED shape shows that IRS-48 consists of -- at least -- two main disk-components: an outer and extended disk beyond \ap55 AU radius, and a closer one, that emits most of the PAH light, as found by~\citet{Maaskant14}. Our VIS2 data does not constrain well the outer disk which mostly lies outside its field of view. Hence, we systematically adopt the simplest axi-symmetrical fitting solution consistent with the literature for this outer-disk, while we focus our fitting effort on the inner parts.

Similarly to the \textit{old} model, we adopt a three disk-components model: two dusty inner- and outer-disks and an additional disk composed of Very Small Particles (VSP) in between. For each of them, the inner radius, outer radius and total dust mass are free parameters. As an initial step, the parameters ``scale height'' and ``flaring exponent'' are shared between all three disks. Other less-impacting parameters such as the surface density exponent, the grain minimum and maximum sizes, the ratio between carbonaceous~\citep[from ][]{Li97} and silicate~\citep[from ][]{Draine84} grains, the ratio between ionized and neutral PAH and the grain-size power-law index or the disk inclination are left-aside for the initial fitting exercise and set to standard values from IRS-48 and disk literature.
The flaring exponent=0.67 \citep{vanderMarel13}, the inclination=50$^{\circ}$ \citep{Geers07,vanderMarel13,Bruderer14}, the surface density exponent=-1 \citep{Andrews13}, the inner-rim radius=55 AU \citep{Geers07}, the silicate over carbonaceous grains mass-ratio=70/30 (prescriptions span 85/15 in~\citet{Wood02}, 45/55 in~\citet{Kim94}; 80/20 in~\citet{Maaskant14}) and the min/max grain sizes=[0.03,4000] $\mu$m \citep{vanderMarel13} of the outer-disk are set parameters, see Table~\ref{table:mcfost_model}.

The effective temperature of the central star is set to 9250 K, mid-way between~\citet{Follette15} work (9500 K) and~\citet{Brown12a} work (9000 K), who also estimated its spectral type to be A0$_{-1}^{+4}$ (9000$\pm$550 K), see Table~\ref{table:star_parameters}.

\subsection{New model: Inner-most Region and Star}

VIS2 data gives stringent constraints on both the flux ratio between the unresolved flux (inner-most disk and star) and resolved flux (further than \ap9 AU). To correct for the resolved-flux depletion of about a factor of 2.7 compared to the unresolved-flux, one can either 1) decrease the inner-most disk opacity to decrease the unresolved flux or 2) increase the PAH disk opacity further than 9 AU to increase the resolved flux. This later solution -- increasing PAH emission in Lp-band (3.8 $\mu$m) and Mp-band (4.78 $\mu$m) -- would immediately lead to a dramatic over-brightness of the main PAH emission features between 4 and 15 $\mu$m, with no (or few) degrees of freedom to compensate for it (i.e. structural parameters such as flaring exponent, scale height, etc).

Decreasing the disk brightness within the resolution angle of the instrument would automatically lead to an under-fitting of the NIR flux. This can easily be corrected by increasing the brightness of the star, de facto decreasing the amount of NIR excess arising from dust needed in the fit. Indeed, as discussed by~\citet{Follette15}, a hotter and especially larger star questions the existence of a third inner-most disk in the \ap1 AU area from the star where this disk was usually needed to account for the 1-3 $\mu$m IR excess. We hence decided to release fitting constraints by allowing the stellar luminosity to vary between 10 and 100 L$_{\odot}$. This implied adding two additional free parameters to tune the extinction of the target in order to fit the photometric fluxes measured in the visible and NIR (R$_v$ taken in $[4.0-10.0]$ and A$_v$ taken in $[10.0-15.0]$).

\subsection{New model: Very Small Particles Disk}

A morphological fit of a simple two-dimensional Gaussian shows that the characteristic location of the resolved flux should lie in a region between 10 and 20 AU. The PAH disk is a natural candidate to account for this flux, since previous work showed that the cavity up to \ap55 AU was not depleted of such grains \citep{Geers07,Maaskant14}. This however means that the bright inner rim of the PAH disk must be fully resolved, i.e. further than \ap7 to 10 AU, in order to account favorably to the contrast ratio between unresolved and resolved flux.

The MIR SED does not show any bump peeking at 9.5 and 18.3 $\mu$m, characteristic of silicate grains smaller than \ap2 $\mu$m. It does instead show a clear deep at 9.8 $\mu$m, de facto eliminating any major silicate grains contribution to the MIR flux; refer to Section~\ref{section:SIL_VSP} for an study on classical thermal grains in the inner-regions. Hence, we decide not to incorporate any small or classical grains in the cavity between the inner- and outer-disks.

To model the very small particles, we follow the prescription of~\citet{Natta93} and use a mixture of carbonaceous VSG and PAH with a ionized/neutral fraction parameter, from~\citet{Li01,Weingartner01,Draine07}. Following the classification of~\citet{Tielens08} (Table 2), we set $a_{min}$=4 \AA~and $a_{max}$=10 \AA~for PAH (ionized and neutral), $a_{min}$=10 \AA~and $a_{max}$=30 \AA~for VSG (referred as PAH clusters and VSG in their table). The grain size power-law index for PAH and VSG is set to -4; only the relative abundances of these three species are left as free parameters for fitting. Finally, we assume that these VSP are quantum heated by UV stellar-flux, i.e. outside local thermodynamic equilibrium (LTE). The quantum heating routines of MCFOST were benchmarked and validated by~\citet{Camps15}.

\section{Resulting Structure}\label{section:diskstruct}

\begin{figure}
  \centering
  \includegraphics[width=\hsize]{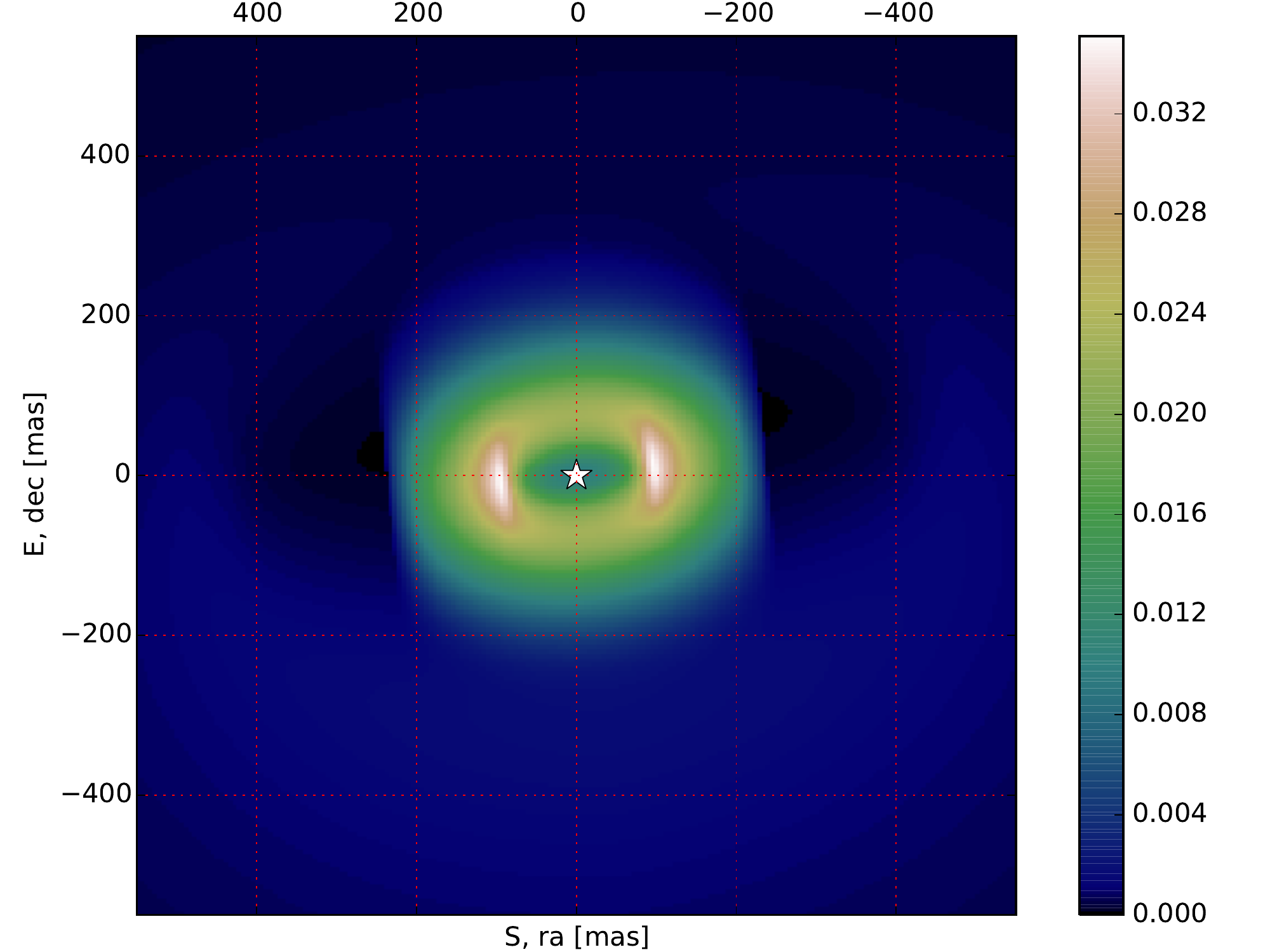}
  \caption{Lp-band model-image normalized to the stellar-flux, obtained from the radiative transfer on our model. This image was obtained out of 15 sub-images generated at wavelengths linearly spread in the Lp-bandpass. The inner-disk between 11 and 26 AU appears much brighter in Lp-band than the outer-disk from 55 AU. Two cavities appear, one inside 11 AU and another one between 26 and 55 AU.}
  \label{figure:img_good}
\end{figure}

This best solution is obtained for a scenario where the innermost disk (0.4-1 AU) has such a low mass that it does not contribute to IRS-48 flux in any wavelength. The new IRS-48 model is hence composed of two disk-elements: a VSP-only disk between 11 and 26 AU and an outer-disk from 55 AU, as portrayed in the Lp-band model-image (see Figure~\ref{figure:img_good}). The star has a higher luminosity $L_*$=42 L$_{\odot}$, leading to a larger radius $R_*$=2.5 R$_{\odot}$ than previously reported. The main parameters of the solution are listed in Table~\ref{table:mcfost_model}. Note again that the VIS2 data is mostly blind to the outer-disk; the solution found represents the simplest centro-symmetrical structure that satisfied the SED fluxes.

\subsection{General Results}

\begin{table}
\caption{Model parameters of the radiative transfer of IRS-48, comparison between the best-fit for the new model and the \textit{old} model as a merge of~\citet{Bruderer14} and~\citet{Maaskant14}.}
\begin{center}
\begin{tabular}{ l c c }
  \hline\hline
  Parameter & \textit{old} model & New model \\ \hline
  
  \multicolumn{3}{c}{Stellar parameters} \\
  Temperature & 9250 K & 9250 K* \\
  Luminosity & 14.3 L$_{\odot}$ & 42 L$_{\odot}$ \\
  Distance & 120$pc$ & 120$pc$* \\
  R$_v$ & 5.5 & 6.5 \\
  A$_v$ & 11.5 & 12.9 \\ \hline
  
  \multicolumn{3}{c}{Global disks parameters} \\
  Inclination & 50$^{\circ}$ & 50$^{\circ}$ * \\
  PA & 96$^{\circ}$ & 96$^{\circ}$ * \\
  Scale height & 10 AU & 14 AU \\
  Ref. radius & 60 AU & 60 AU* \\ \hline
  
  \multicolumn{3}{c}{Inner-most disk} \\
  $R_{in}$ Inner & 0.4 AU & -- \\
  $R_{out}$ Inner & 1 AU& -- \\
  Flaring exp. $\beta$ & 1.3 & -- \\
  Surface density exp $p$ & -1 & -- \\
  Grain-size power-law & -3.5 & -- \\
  Dust Mass & 8.0\pow{-12} M$_{\odot}$& -- \\
  Silicate grains Mass & 1\% & -- \\
  Carb. grains Mass & 99\% & -- \\
  Grain sizes $a$ & 0.03 - 30 $\mu$m & -- \\ \hline
  
  \multicolumn{3}{c}{VSP-ring} \\
  $R_{in}$ Inner & 1 AU & 11 AU \\
  $R_{out}$ Inner & 50 AU& 26 AU \\
  Flaring exp. $\beta$ & 1.3 & 0.6 \\
  Surface density exp $p$ & -1 & -0.1 \\
  Grain-size power-law & -- & -4* \\
  Dust Mass & 8.0\pow{-10} M$_{\odot}$ & 3.7\pow{-10} M$_{\odot}$ \\
  VSG Mass & 0\% & 20\% \\
  VSG sizes & -- & 10 - 30 \AA~* \\
  Neutral PAH Mass & 50\% & 60$\%$ \\
  Ionized PAH Mass & 50\% & 20$\%$ \\
  PAH sizes & 5 \AA~& 4 - 10 \AA~* \\ \hline
  
  \multicolumn{3}{c}{Outer disk} \\
  $R_{in}$ & 55 AU & 55 AU* \\ 
  $R_{out}$ & 160 AU& 250 AU \\
  Flaring exp. $\beta$ & 1.3 & 0.67* \\
  Surface density exp $p$ & -1 & -1* \\
  Grain-size power-law & -3.5 & -3.5* \\
  Dust Mass & 1.0\pow{-5} M$_{\odot}$ & 9.0\pow{-6} M$_{\odot}$ \\
  Silicate grains Mass & 70$\%$ & 70$\%$* \\
  Carb. grains Mass & 30$\%$ & 30$\%$* \\
  Grain sizes $a$ & 0.03 - 4000 $\mu$m & 0.03 - 4000 $\mu$m* \\
\end{tabular}\par
\end{center}

*=set-value parameter
\label{table:mcfost_model}
\end{table}

\begin{figure}[!]
\centering
  \includegraphics[width=\hsize]{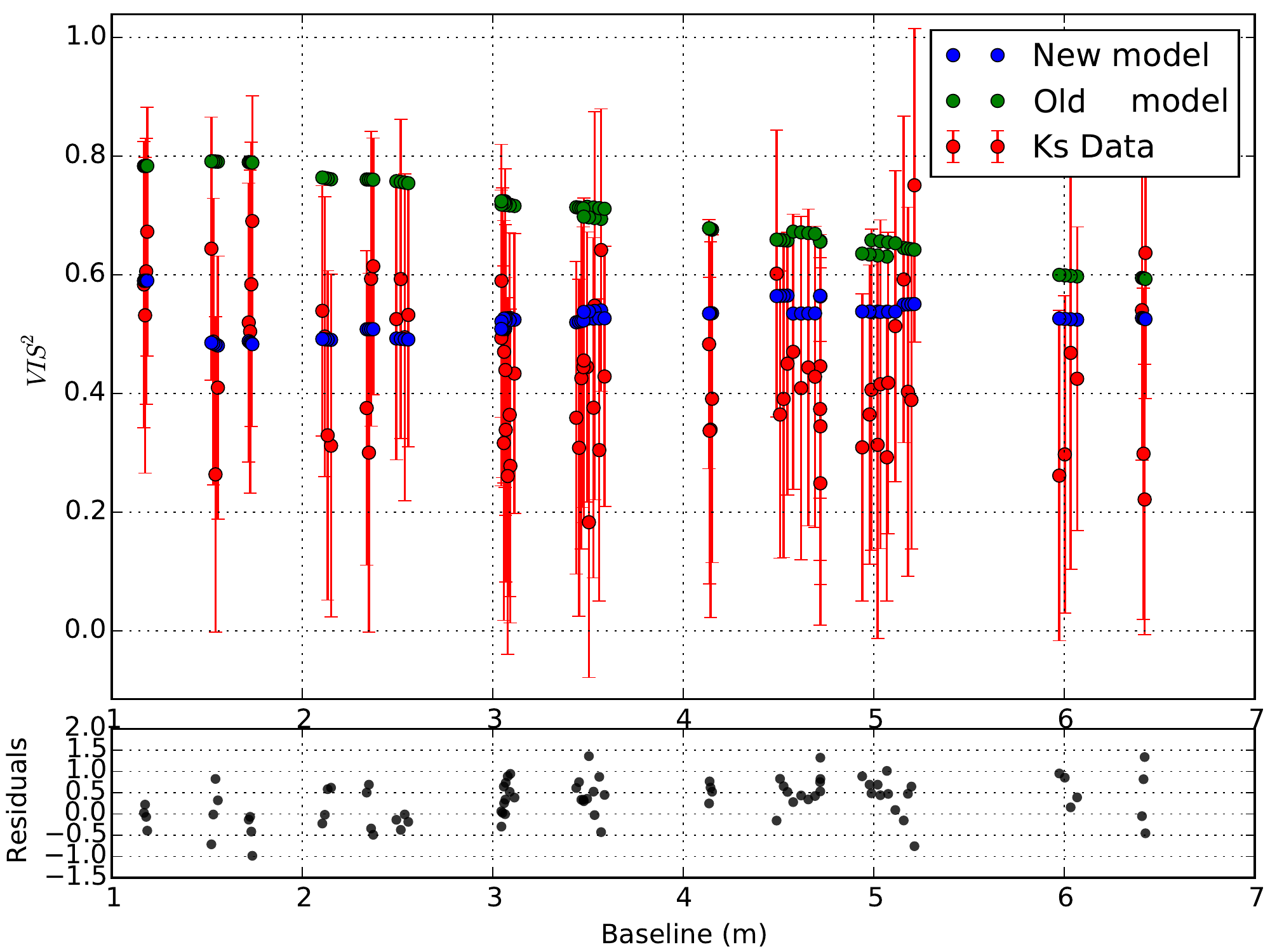}\par%
  \includegraphics[width=\hsize]{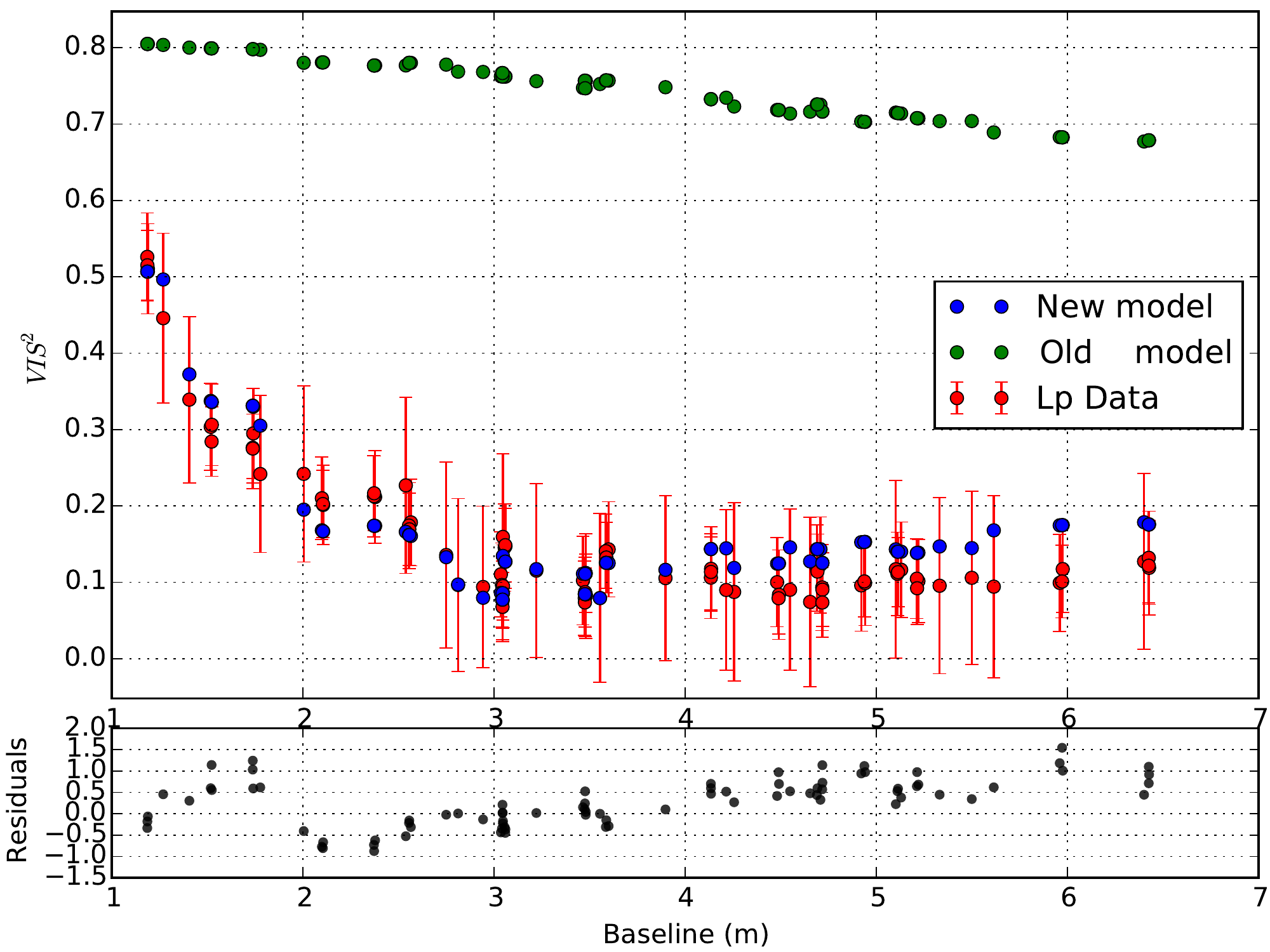}\par%
  \includegraphics[width=\hsize]{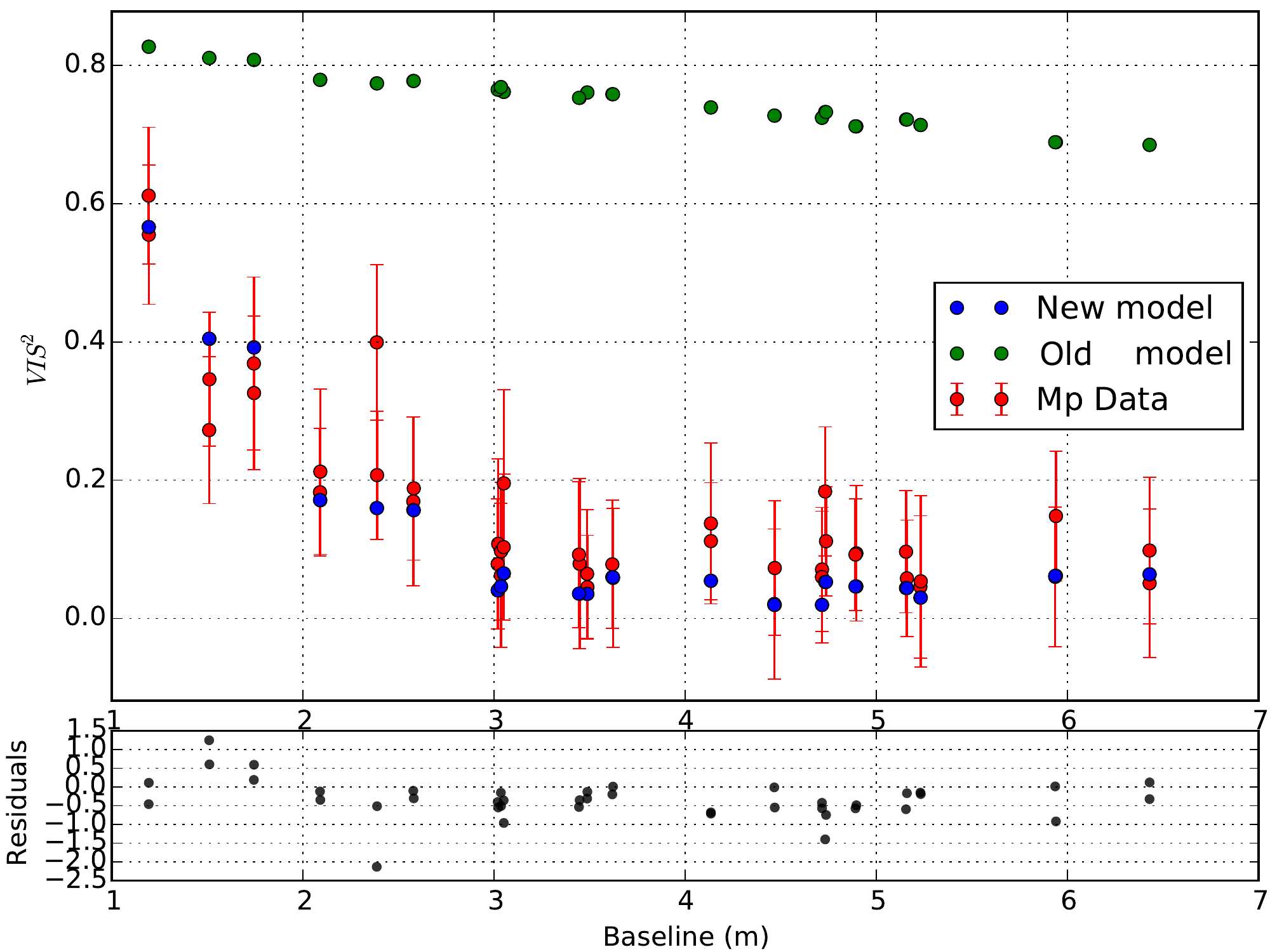}
\caption{Diagram showing the Ks-band (top), Lp-band (middle) and Mp-band (bottom) VIS2 data at all epochs, the calculated squared-visibility from the \textit{old} model-image (green) and those of our model (blue) as a function of the un-projected baselines in meter. Most residuals for our model span $\pm1$ while they exceed +4 for the \textit{old} model. Although the Ks-band data is much noisier, our model still does provide an improvement over the \textit{old} model.}
\label{figure:VIS2_bl}
\end{figure}

\begin{figure*}
  \centering
  \includegraphics[width=\hsize]{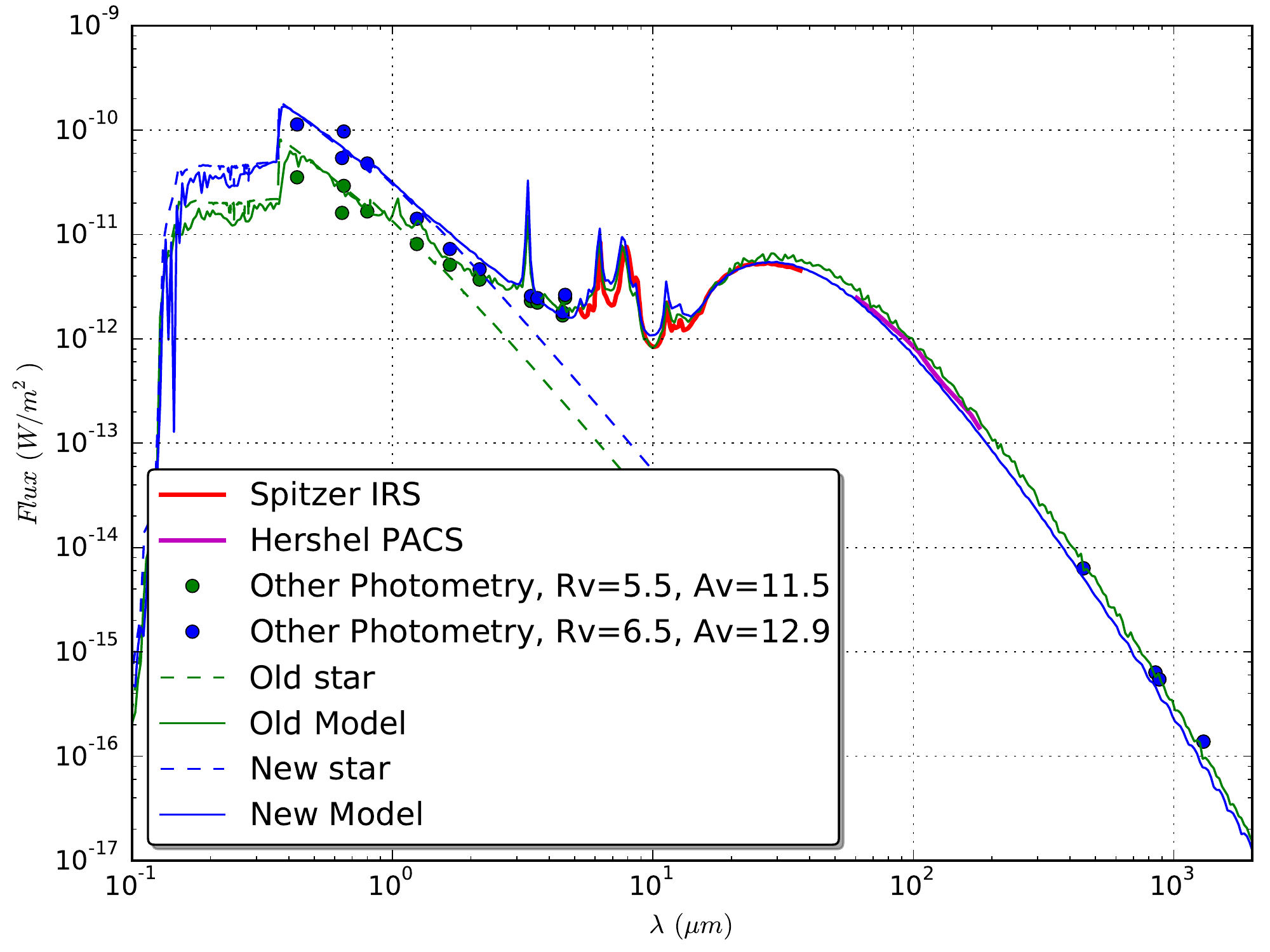}
  \caption{SED of IRS-48 obtained from the \textit{old} model -- or literature model -- (green line) and our model (blue line). Both explain the photosphere of the star (although with different extinction parameters), the NIR-excess and the outer-disk SED-bump centered on 20 $\mu$m.}
  \label{figure:sed_good}
\end{figure*}

Figure~\ref{figure:sed_good} shows the SED for both models with their respective interstellar extinction. Our new model provides the same fit quality as the \textit{old} model. It matches very closely all the main features of the SED: the photosphere of the star, the NIR-excess, the PAH emission-features and the outer-disk. Most importantly, it improves dramatically the fit on the VIS2 data, which it explains entirely: Figures~\ref{figure:VIS2_bl} show the comparison between the VIS2 data in Ks-, Lp- and Mp-band, and both models.

\subsection{Central Star}

The solution of a $L_*$=42 L$_{\odot}$ star with a 3.7\pow{-10} M$_{\odot}$ mass VSP-ring and no inner-most disk (0.4-1 AU) is a complex and precise balance between the fit to the SED and to the VIS2 data. Indeed, the NIR SED requires either a brighter star than the \textit{old} model, or a similar star with an inner-most disk to account for a large NIR-excess. The VIS2 data however pinpoints a precise contrast ratio between the resolved VSP-ring and the unresolved star and inner-most disk altogether. Adding even the smallest inner-most disk-mass creates an excessive NIR emission which must be compensated with a large increase of the VSP disk-mass to fit the VIS2 data. Such an increase however dramatically over-estimates the PAH emission features between 5.5 $\mu$m and 18 $\mu$m. This strongly pushes to a solution with no inner-most disk inside the VSP-ring.


We follow the isochrones of~\citet{Siess00} Young Stellar Object (YSO) evolutionary models with a luminosity increased from 14 to 42 L$_{\odot}$ and an effective temperature of 9250 K to find a larger stellar radius $R_*$=2.5 R$_{\odot}$, rather than 1.8 or 1.4 R$_{\odot}$ previously reported. The main stellar parameters are listed in Table~\ref{table:star_parameters}, with a comparison to the previous literature values.

\begin{table}
\caption{New stellar model for $\rho$ Oph IRS-48 central star.}
\begin{center}
\begin{tabular}{ l c c c}
  \hline\hline
   & This work & F15 & B12 \\
  \hline
  T (K) & 9250 & 9500 & 9000\\
  R (R$_{\odot}$) & 2.5$^{(1)}$ & 1.8 & 1.4 \\
  L (L$_{\odot}$) & 42 & 23.6 & 14.3 \\
  Age (Myr) & 4.1$^{(1)}$ & 8 & $>$15 \\
  M (M$_{\odot}$) & 2.5$^{(1)}$ & 2.2 & 2.0 \\
  A$_v$ & 12.9 & 12.0 & 11.5 \\
  R$_v$ & 6.5 & (4.0) & 5.5 \\
  A$_k$ & (1.8) & 1.5 & (1.6) \\
  \hline
  \end{tabular}\par
\end{center}
$^{(1)}$ according to~\citet{Siess00} YSO evolutionary models.\\
F15$=$\citet{Follette15}, B12$=$\citet{Brown12a}
  \label{table:star_parameters}
\end{table}

To compensate for the excess stellar luminosity, interstellar extinction was re-evaluated to A$_v$=12.9 and R$_v$=6.5. IRS-48 is 1.4 V-magnitude more extinct, with a more gray R$_v$ than the extinction inferred in~\citet{Brown12a} (A$_v$=11.5 and R$_v$=5.5). This confirms and reinforces the presence of substantial large dust grains in the line of sight \citep{Kim94,Indebetouw05}, typical of star-forming regions such as Ophiuchus. An equivalent value of A$_k$=1.8 is consistent with that of \ap1.6 read on the low spatial-resolution extinction map computed by~\citet{Lombardi08}; their estimate for that Ophiuchus-core region suffers however from a high discrepancy, most probably due to the unknown depth of their targets into this dense molecular cloud.

The radiative modelling shows that there is no extinction due to the circumstellar disk; its \ap50$^{\circ}$ inclination is not large enough to extenuate the stellar flux like (nearly) edge-on disks. All extinction is due to interstellar extinction, most of which arises from the local molecular cloud of $\rho$ Oph.

The evolutionary tracks imply that such a star falls onto the \ap4 million years locus, with $M_*$=2.5 M$_{\odot}$. This is much younger than the 8 or 15 Myr previously inferred, partly solving the evolutionary puzzle on the presence of a disk. A discussion on the age of IRS-48 is to be found in Section~\ref{section:age}.

\subsection{Very Small Particles ring}\label{section:gory_details}

\begin{figure}
  \centering
  \includegraphics[width=\hsize]{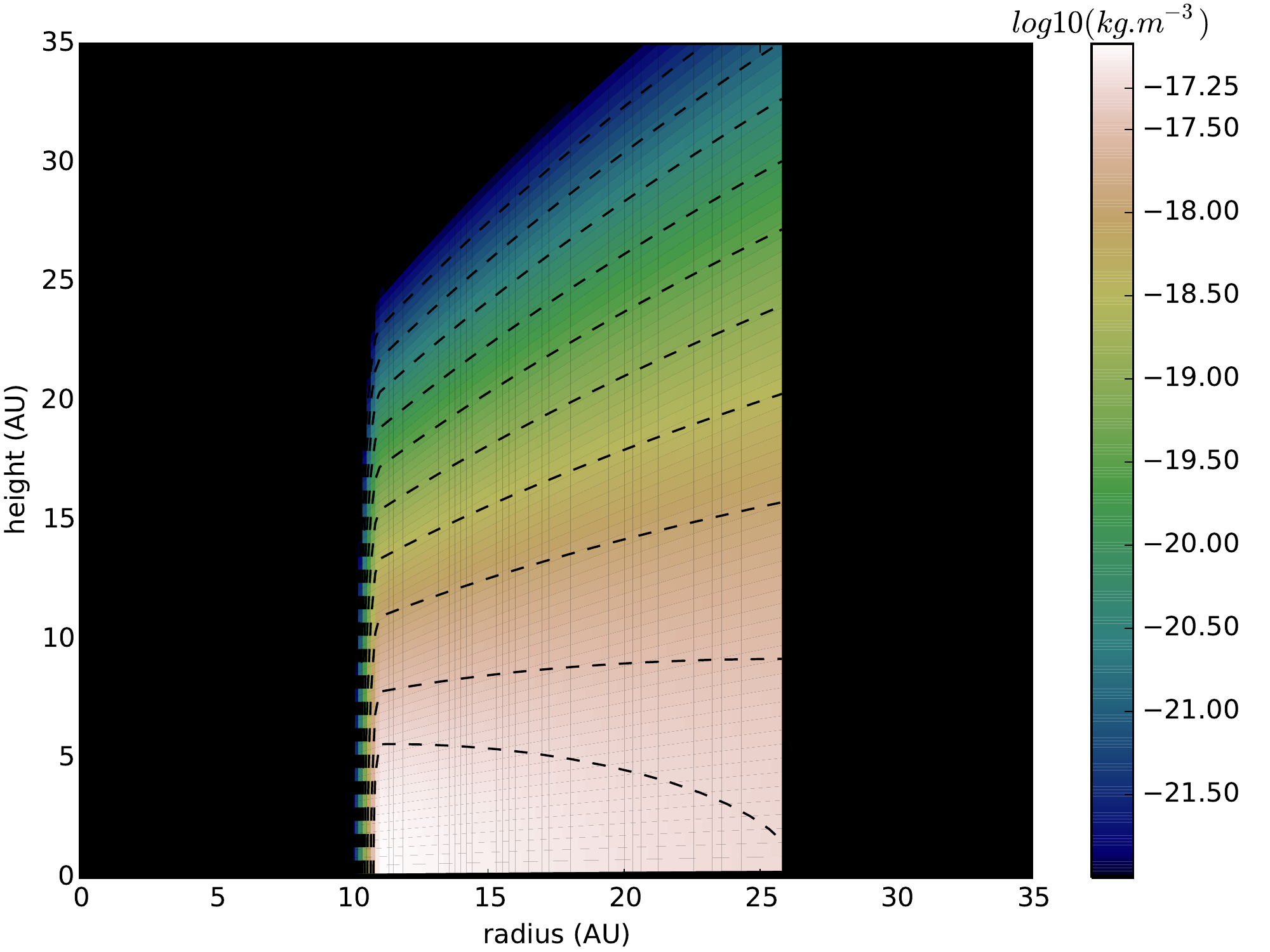}
  \caption{Map showing the PAH dust density [kg.m$^{-3}$] in log10 scale, in a cut perpendicular to the plane of the disk. Iso-contours on the map are shown at the levels of the color-bar ticks. The densest region of the VSP-ring spans from 11 to 20 AU with an average height of \ap5 AU above the mid-plane. The ring shows a smooth and nearly flat density structure with increasing radius due to the high surface density exponent of -0.1.}
  \label{figure:PAH_density_dust}
\end{figure}

This new model consists of a smooth centro-symmetrical and optically thin VSP-only ring between 11 and 26 AU. It is responsible for all of the IR-excess until \ap13 $\mu$m, where the outer disks starts dominating the dust-emission. In the fitting exercise, the VIS2 data constrained mostly the structure of the VSP-ring, while the MIR-SED PAH emission features constrained the relative abundance of the PAH grains and VSG. The VSP density tops at \ap1\pow{-17} kg.m$^{-3}$ (see Figure~\ref{figure:PAH_density_dust}).


\begin{figure*}[!]
\centering
  \includegraphics[width=0.5\hsize]{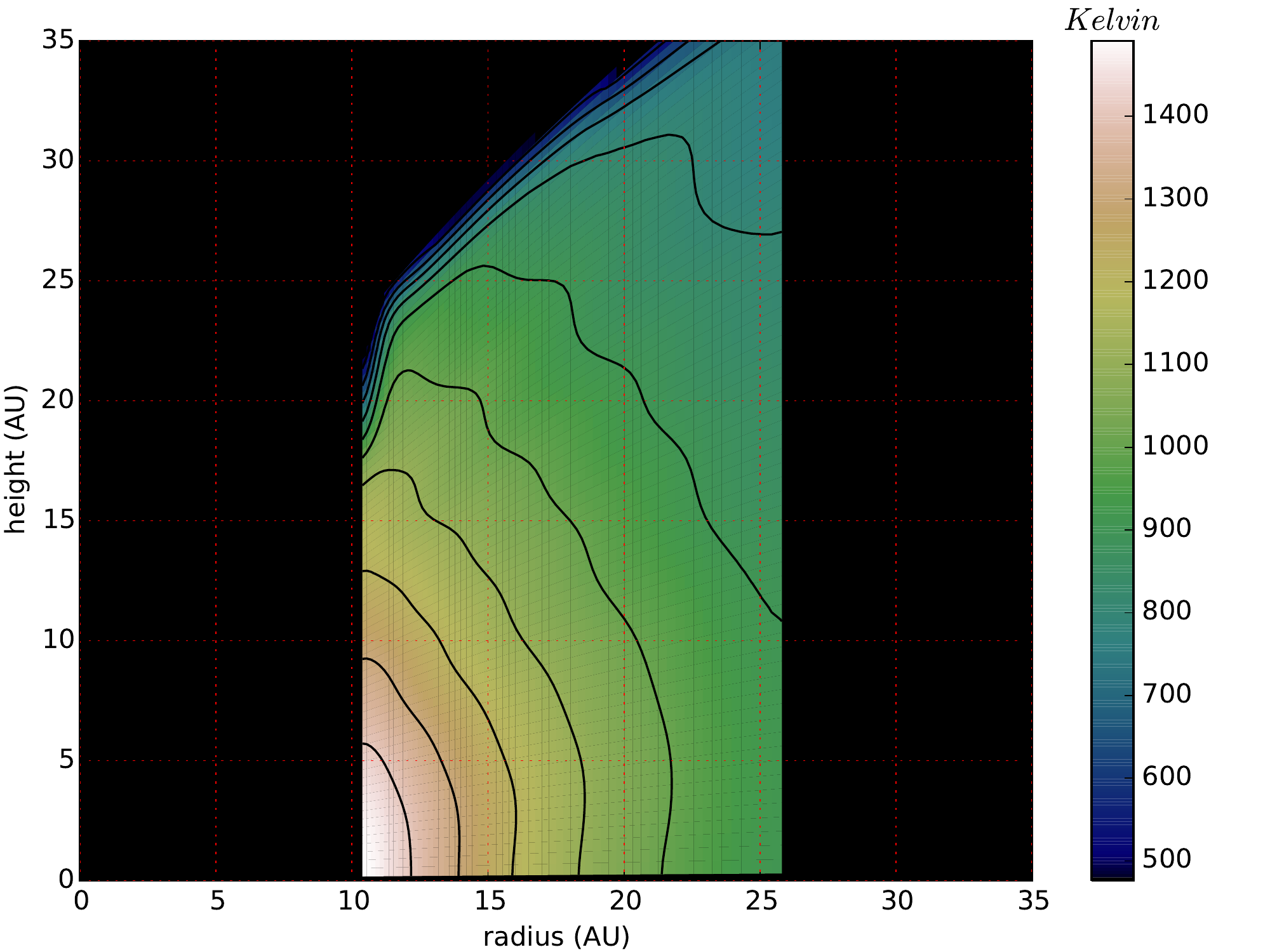}%
  \hfill%
  \includegraphics[width=0.5\hsize]{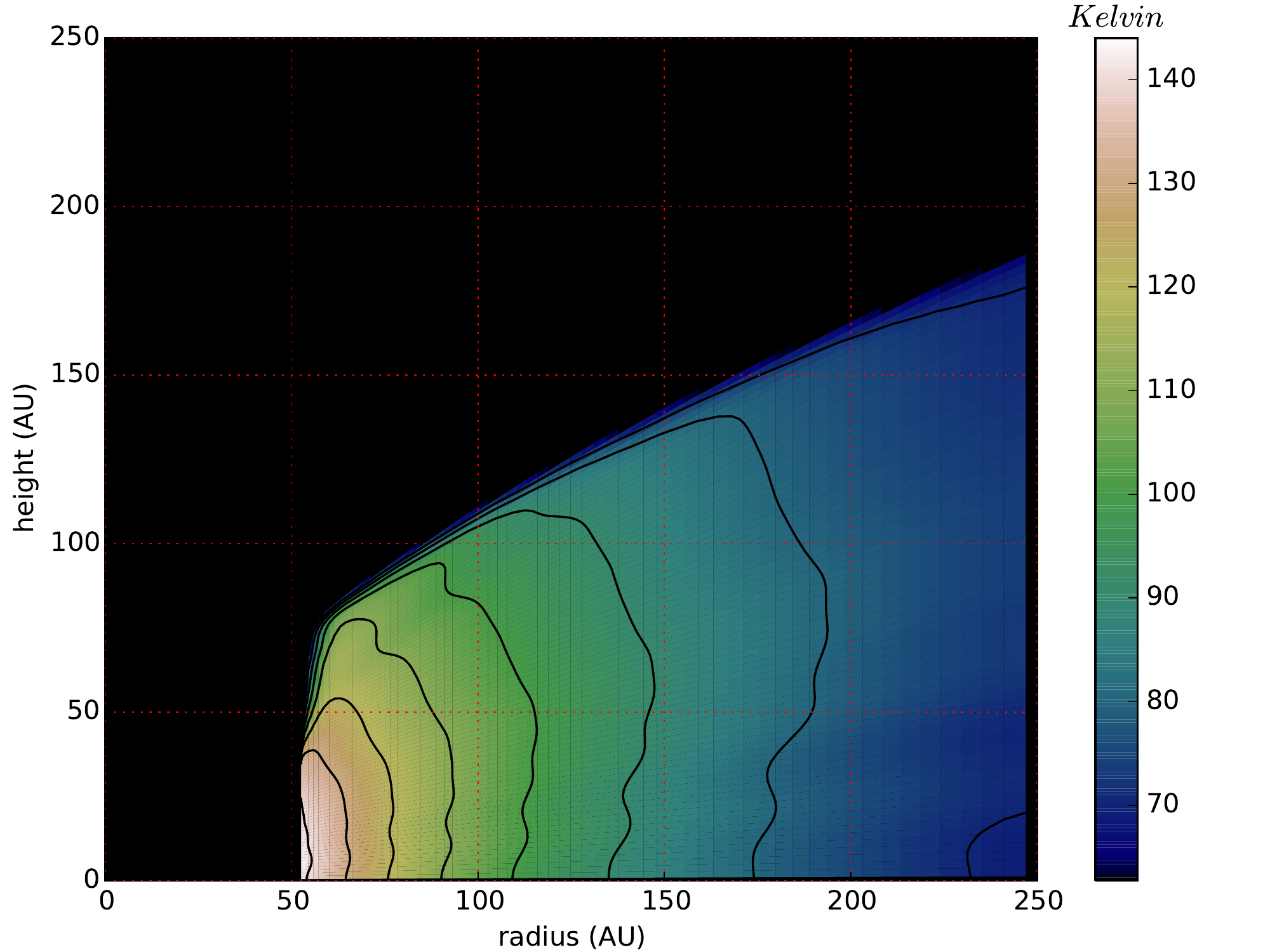}%
\caption{Map showing the temperatures (Kelvin) of the neutral PAH dust (left) and of the outer-disk grains (right), in a cut perpendicular to the plane of the disk. Iso-contours on the map are shown at the levels of the color-bar ticks. Compared to neutral PAH, ionized PAH grains and carbonaceous VSG follow a very similar temperature distribution, but with lower temperatures bounds: T$_{max}$\ap1100 K and T$_{min}$\ap400 K for ionized PAH and T$_{max}$\ap550 K and T$_{min}$\ap250 K for VSG}
\label{figure:temperatures}
\end{figure*}

The resulting disk temperature is shown in Figure~\ref{figure:temperatures} for neutral PAH grains. Because of the close proximity to the star, the time between two successive absorptions is much lower than the relaxation time of the PAH grains. In this case, MCFOST treats quantum heating as ``quasi-equlibrium`` and a temperature can be defined (i.e. low dispersion on the temperature probability distribution). It reaches particularly high values for neutral PAH grains, T$_{max}$\ap1500 K and T$_{min}$\ap600 K. Ionized PAH grains and carbonaceous VSG follow a very similar temperature distribution, with lower temperatures due to their difference in electrical charge, mass and structure: T$_{max}$\ap1100 K and T$_{min}$\ap400 K for ionized PAH and T$_{max}$\ap550 K and T$_{min}$\ap250 K for VSG.

\subsection{Outer Disk}\label{section:outerdisk}

As previously highlighted, the interferometric data is mostly blind to the outer-disk, and is not sensitive in any case to the southern asymmetry in mm-wavelengths observed with ALMA. We fitted the simplest centro-symmetric disk model to satisfy the FIR and mm-wavelength SED.

\begin{figure}
  \centering
  \includegraphics[width=\hsize]{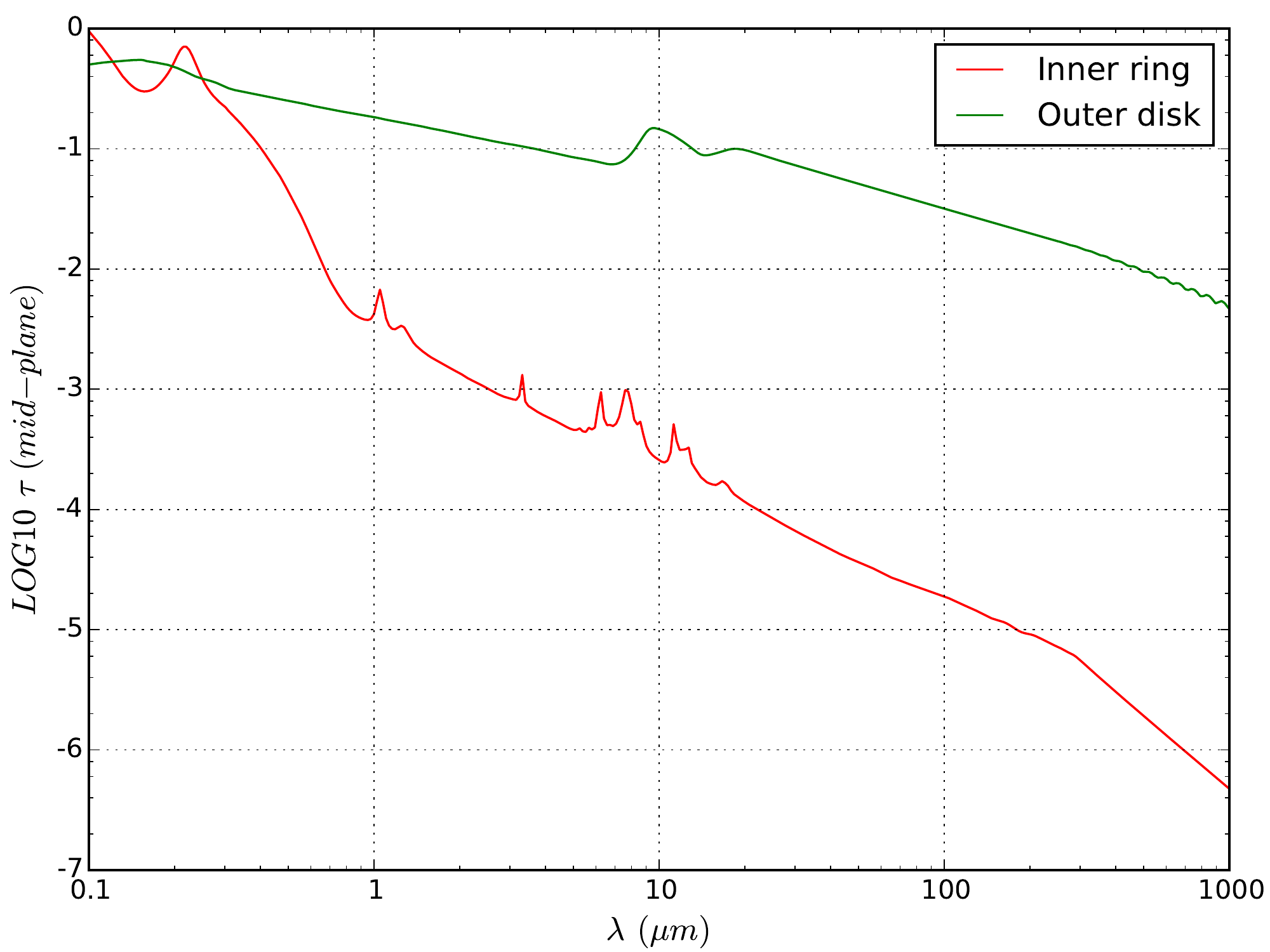}
  \caption{Optical thickness $\tau$ taken in the mid-plane as a function of wavelength for the VSP-ring and the outer-disk. One recognizes the PAH lines from the VSP-ring and the two silicates bumps of the outer-disk at 9.2 and 18 $\mu$m. The VSP-ring becomes extremely thin for wavelengths above a micron.}
  \label{figure:tau_midplane}
\end{figure}

This centro-symmetric approach is however not void of meaning. Indeed, the outer disk is optically thin in its emission window (for wavelengths $\gtrsim$13 $\mu$m, $\tau$ in the mid-plane $<$10$^{-1}$, see Figure~\ref{figure:tau_midplane}). Hence the precise azimuthal location of grains (azimuthally asymmetric or not) only plays a secondary role in fitting the SED.

In order to satisfy the mm-wavelength SED, the outer-radius of the disk is found at \ap250 AU, similarly to~\citet{Maaskant14} (225 AU), although SED fitting is not very sentivite to this parameter. For that disk extend, the total mass of the outer-disk in the MCFOST model is 9\pow{-6} M$_{\odot}$ (3 M$_{Earth}$) for grains between 30nm and 4mm, with a grain-size power law of -3.5. This mass is similar although lower than~\citet{Bruderer14} (1.6\pow{-6} M$_{\odot}$, 5.3 M$_{Earth}$) and 9 M$_{Earth}$~found by~\citet{vanderMarel13} for the same grain-size distribution and bounds. This is due to the fact that the later mass-estimate obtained from 685Ghz ALMA flux was biased as they used a lower temperature, T=60 K at 60 AU, than what is currently observed in our model. Indeed, our star is brighter, hence heats the outer-disk more efficiently: as seen in Figure~\ref{figure:temperatures}, the inner-rim temperature peaks at $T_{dust}$=148 K at 55 AU.

We update their dust mass calculations with the values of our model to find $\tau_{685GHz}$=0.14 for T=140 K at 60 AU, using their expression
  \begin{equation}\label{eqn:taustuff}
     F_{\nu} = \Omega\times B_\nu(T_{dust})\times(1-\exp^{-\tau_\nu}).
  \end{equation}
When we integrate over the whole disk-model the optical thickness at $\nu$=685Ghz ($\lambda$=440 $\mu$m), we find $\tau_{685Ghz}$=0.19, closely consistent with the previous 0.14 found independently using ALMA flux.
Using $M_{dust}=21\times\tau_{685Ghz}\sqrt{a_{max}}$ from~\citet{Draine06,vanderMarel13} under the same hypothesis, we find M$_{dust}$=3 M$_{Earth}$ for $\tau$=0.14 and 4 M$_{Earth}$ for $\tau$=0.19, consistent with the outer-disk dust mass in our radiative transfer model (3 M$_{Earth}$).

\begin{figure*}
  \centering
    \includegraphics[trim={0.9cm 0 1.2cm 0},clip,width=0.33\hsize]{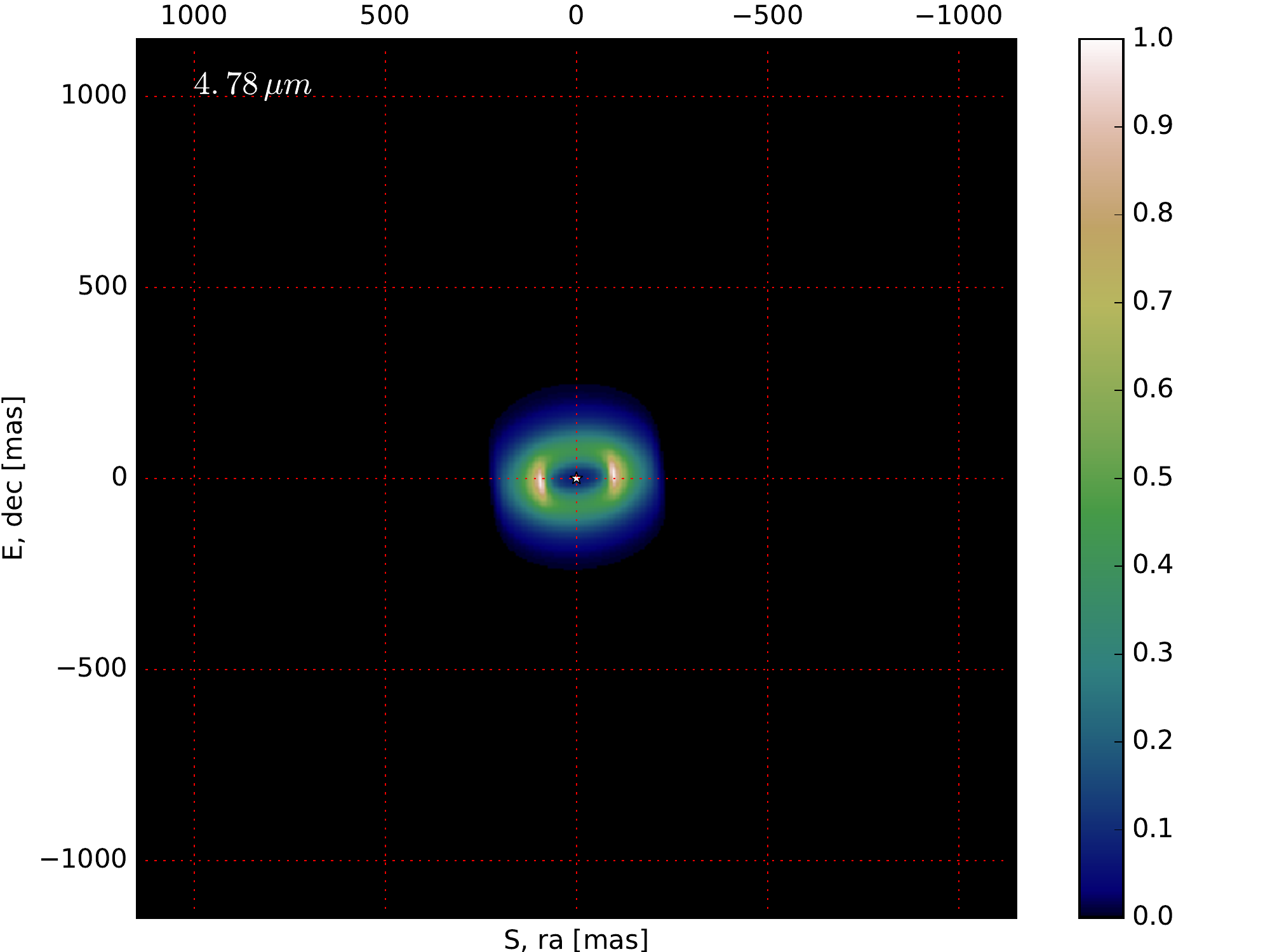}%
    \includegraphics[trim={0.9cm 0 1.2cm 0},clip,width=0.33\hsize]{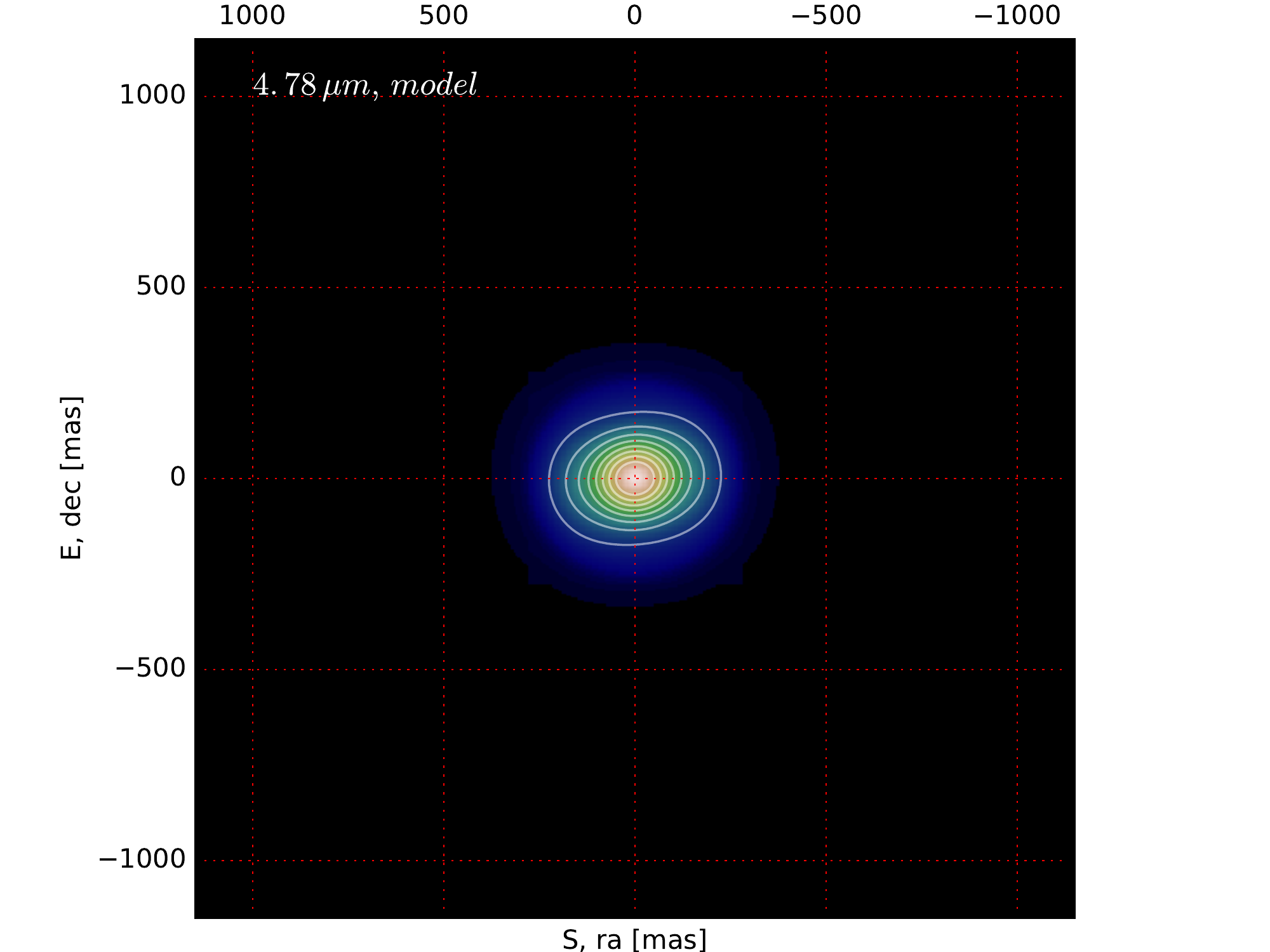}%
    \includegraphics[trim={0.9cm 0 1.2cm 0},clip,width=0.33\hsize]{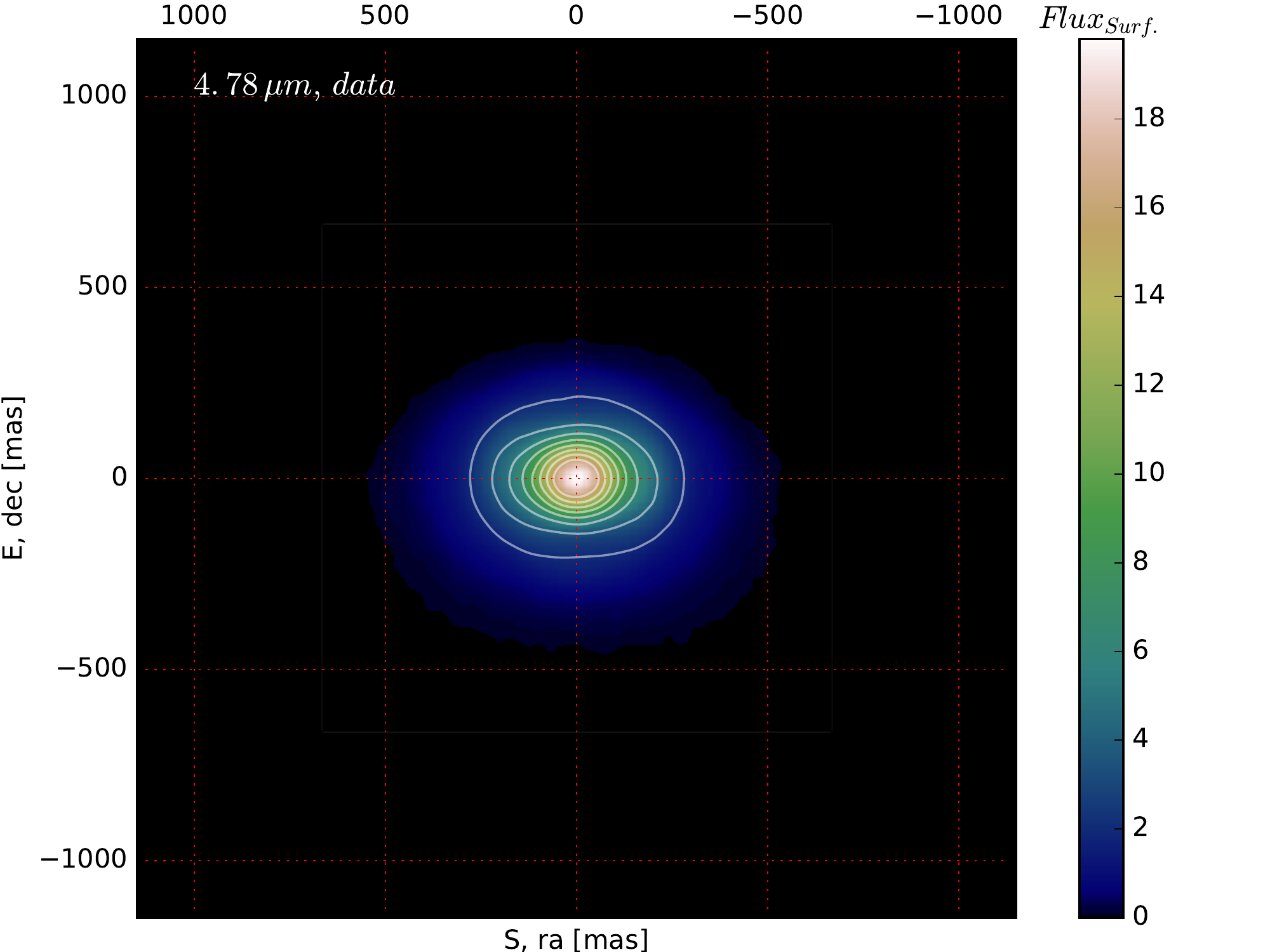}\par
    \includegraphics[trim={0.9cm 0 1.2cm 0},clip,width=0.33\hsize]{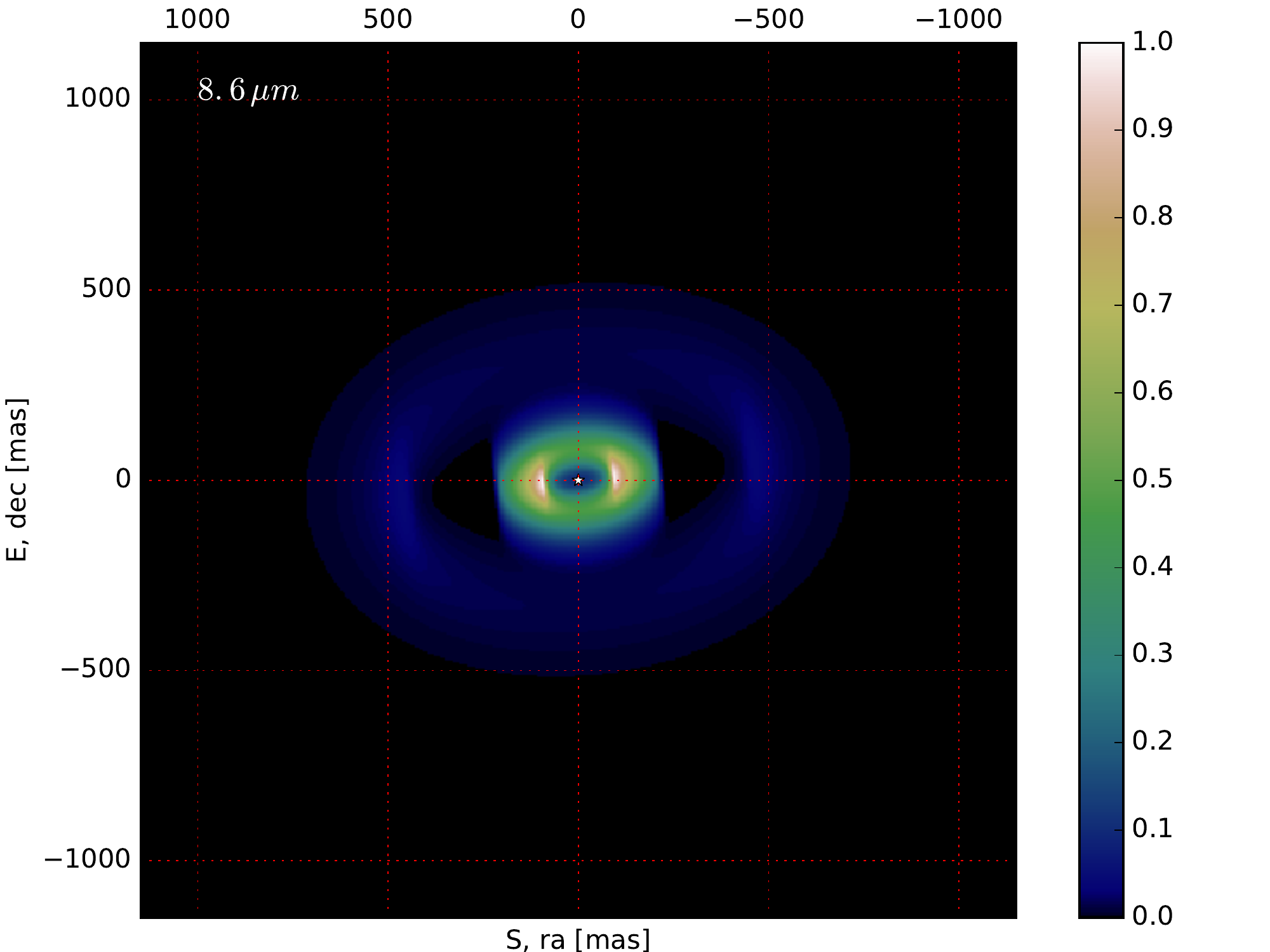}%
    \includegraphics[trim={0.9cm 0 1.2cm 0},clip,width=0.33\hsize]{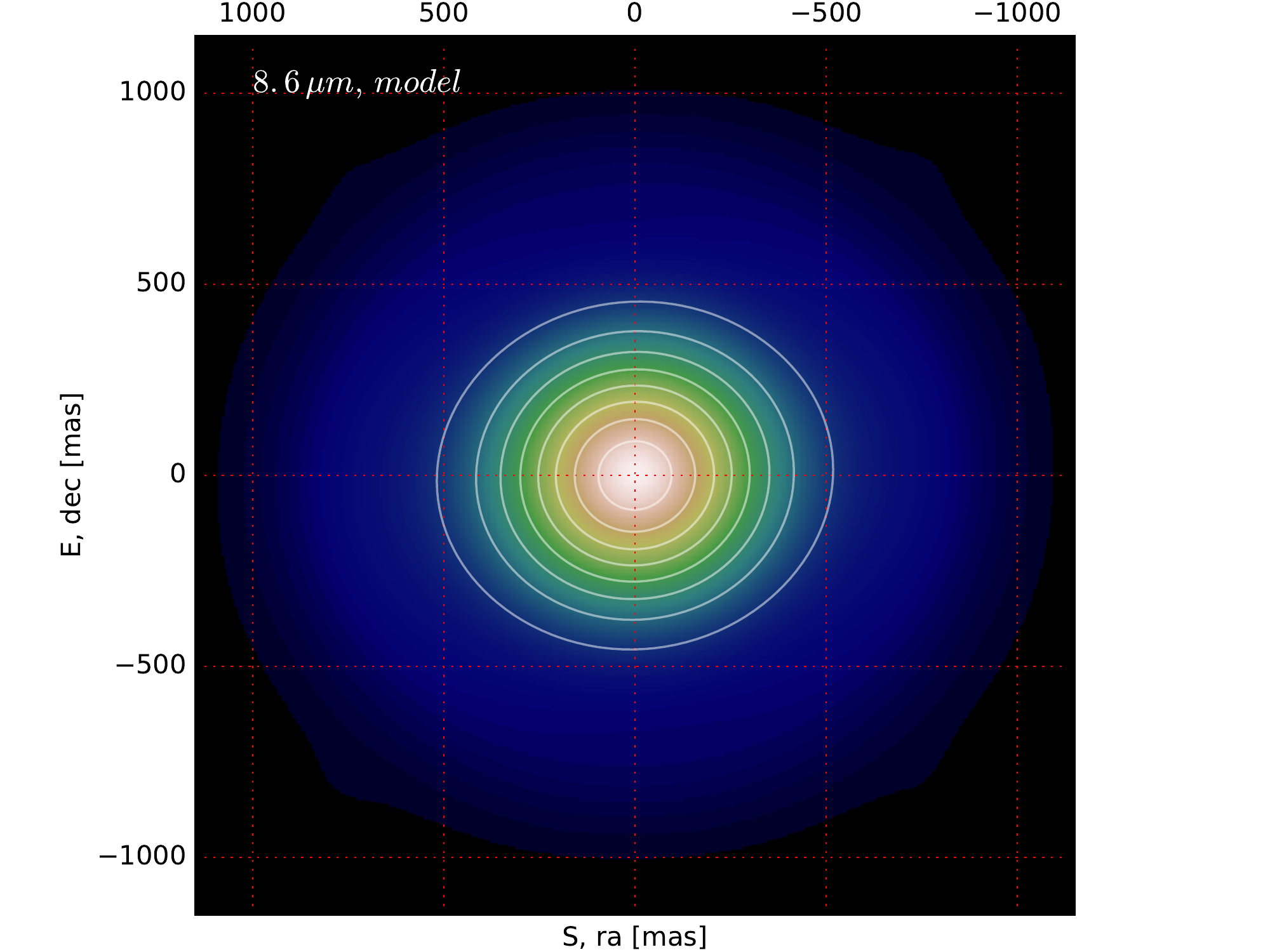}%
    \includegraphics[trim={0.9cm 0 1.2cm 0},clip,width=0.33\hsize]{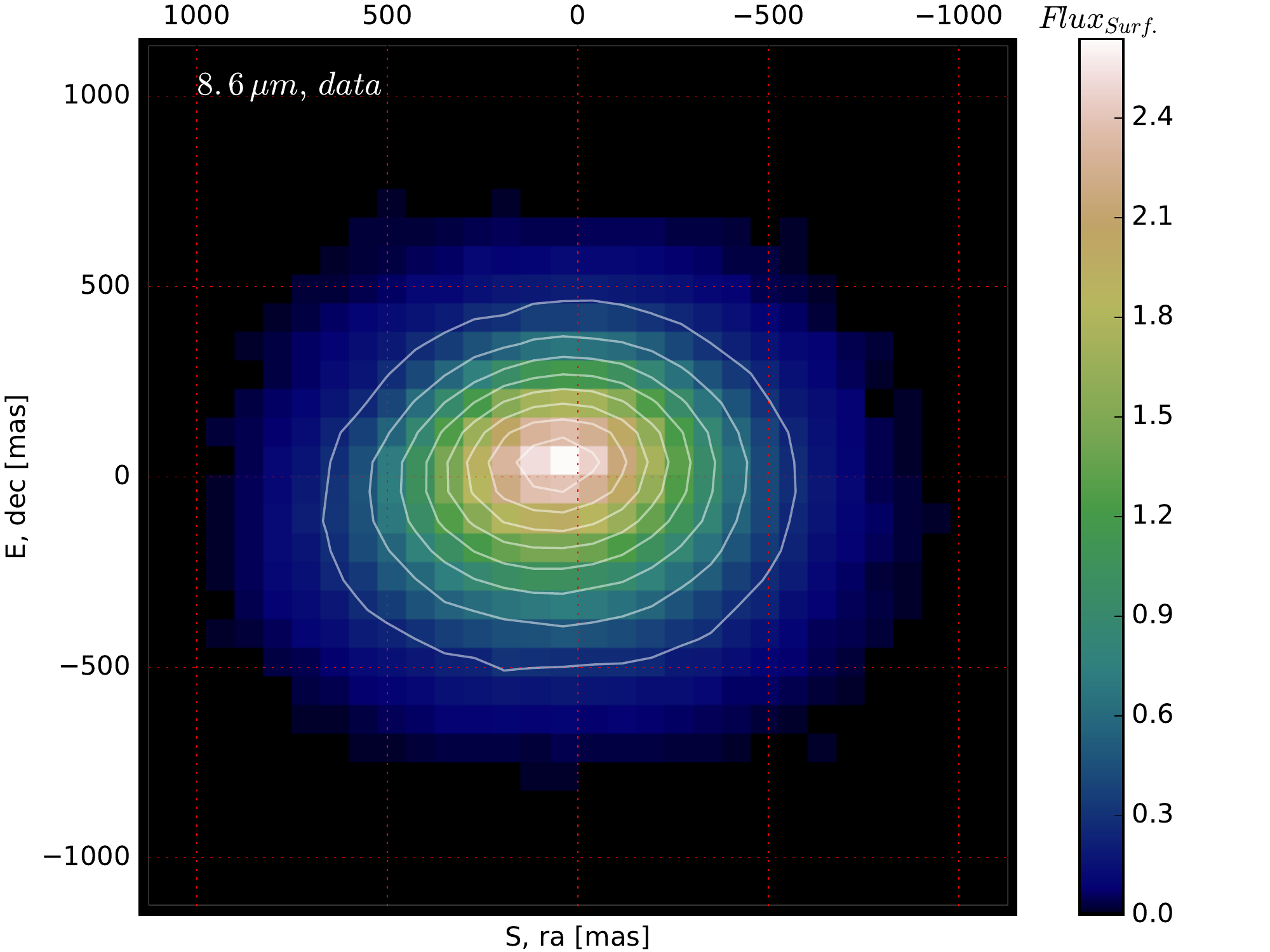}\par
    \includegraphics[trim={0.9cm 0 1.2cm 0},clip,width=0.33\hsize]{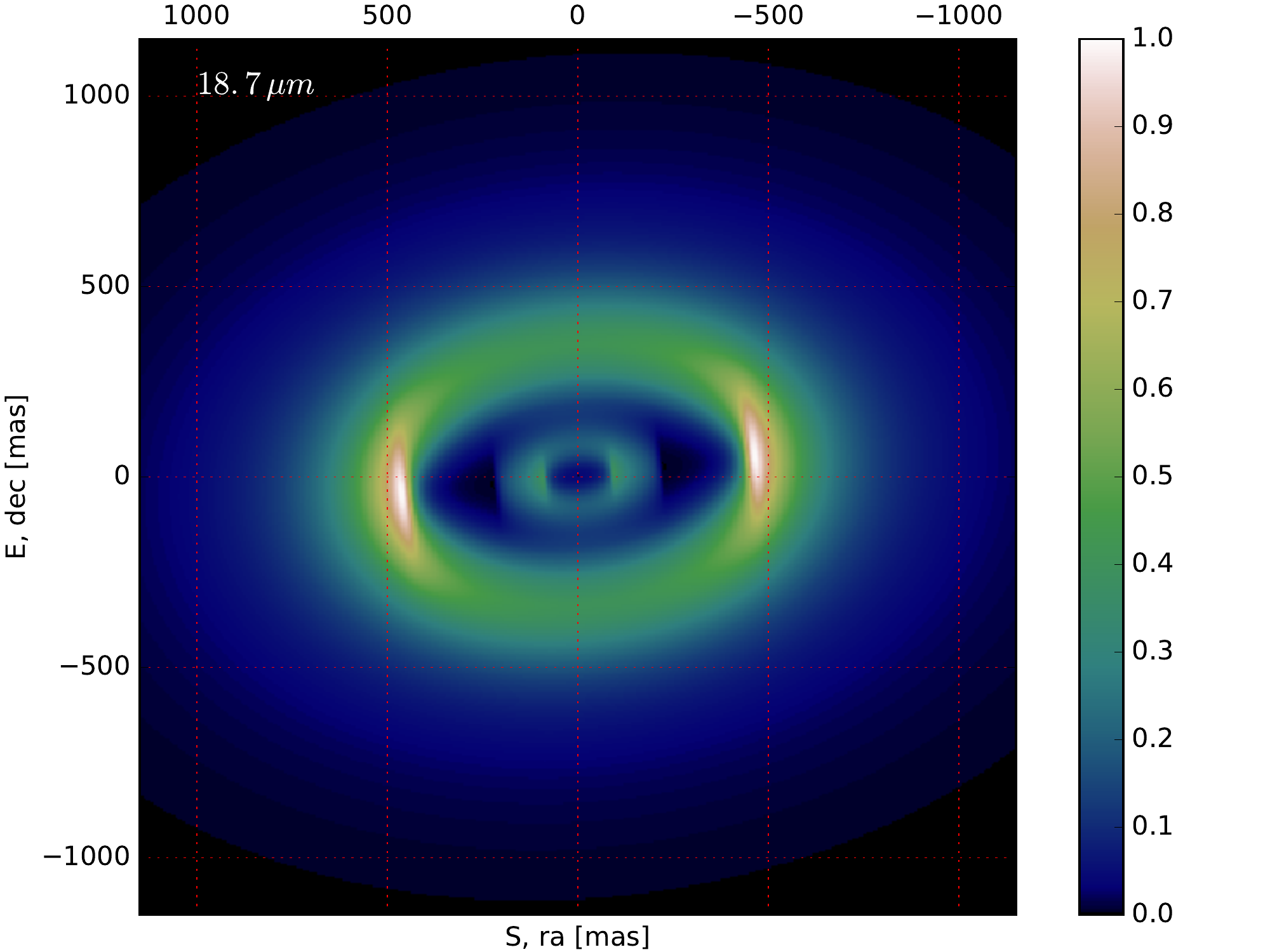}%
    \includegraphics[trim={0.9cm 0 1.2cm 0},clip,width=0.33\hsize]{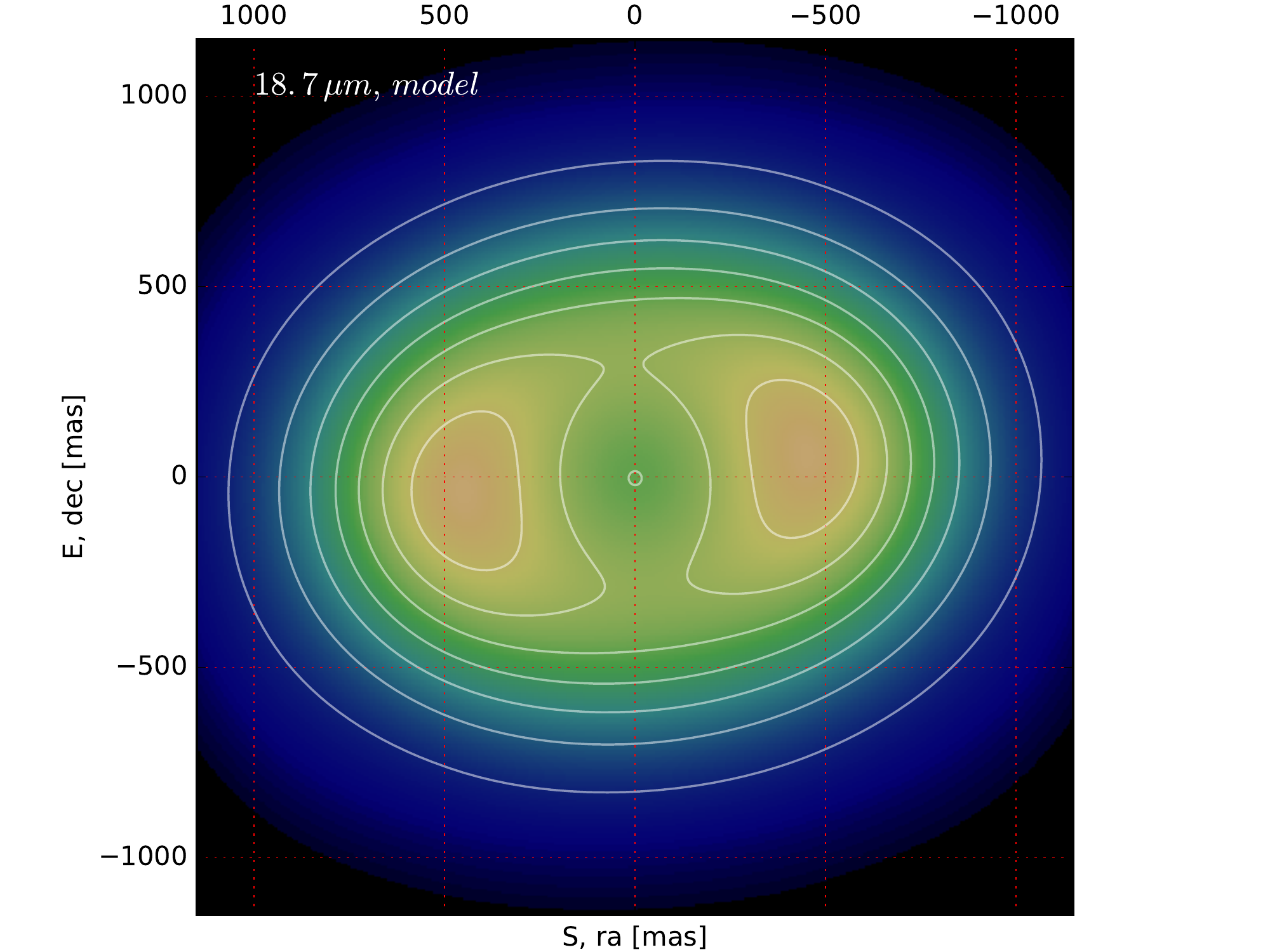}%
    \includegraphics[trim={0.9cm 0 1.2cm 0},clip,width=0.33\hsize]{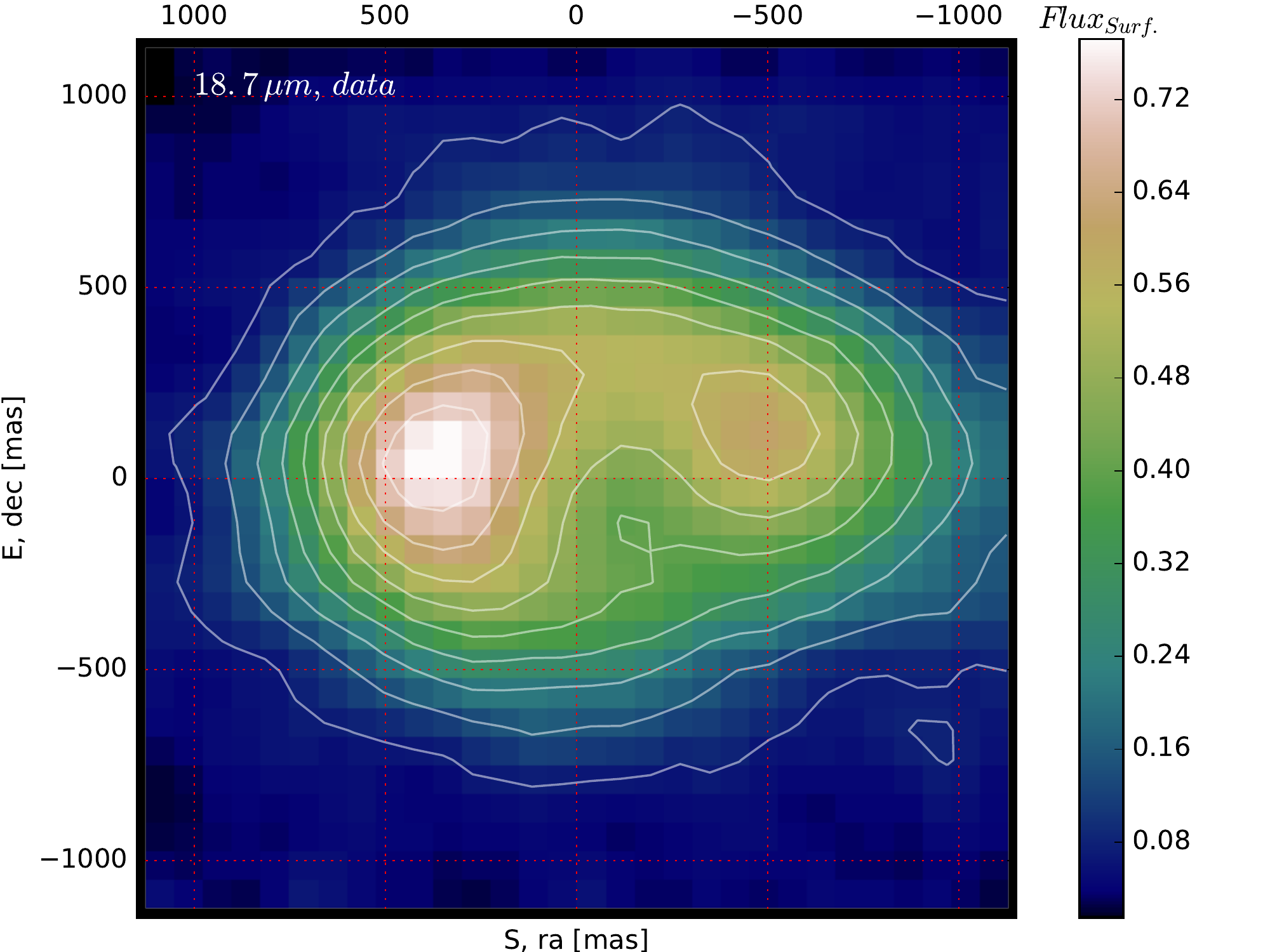}
  \caption{Synthetic-images (left and middle columns) and data-images (right column) at wavelengths 4.78 $\mu$m (Mp-band, top row), 8.6 $\mu$m (middle row) and 18.7 $\mu$m (bottom row). Left-column images are shown in arbitrary flux (the central-star being removed). Synthetic-images in the middle column are convolved with a 8.2m telescope PSF and shown in the color-scale of the observed data-images seen on the right column. Middle and right columns color-scale are fluxes per square-arcsec, normalized to the total flux of the disk, i.e. fraction of the total disk surface brightness. Data-images are taken from this work (Mp-band) and~\citet{Geers07} (8.6, 18.7 $\mu$m). Iso-contours are overlaid at the color-bar tick-values.}
  \label{figure:imgMIR}
\end{figure*}

Figure~\ref{figure:imgMIR} shows model-images generated in the mid-IR with the corresponding imaging data, at 4.78 $\mu$m (Mp-band), 8.6 and 18.7 $\mu$m. Although 8.6 and 18.7 $\mu$m images were not included into the fit -- only SED fluxes were fitted at these wavelengths --, our model can reproduce the observed structure accurately. \citet{Geers07} noted that the 8.6 $\mu$m image is mostly unresolved, with most flux arising from PAH in the inner 30 AU, while the 18.7 $\mu$m image is highly resolved, with most flux arising from the inner rim of the outer-disk located at \ap55 AU.

\subsection{Classical Thermal Grains in the VSP-ring}\label{section:SIL_VSP}

\begin{figure}
    \includegraphics[width=\hsize]{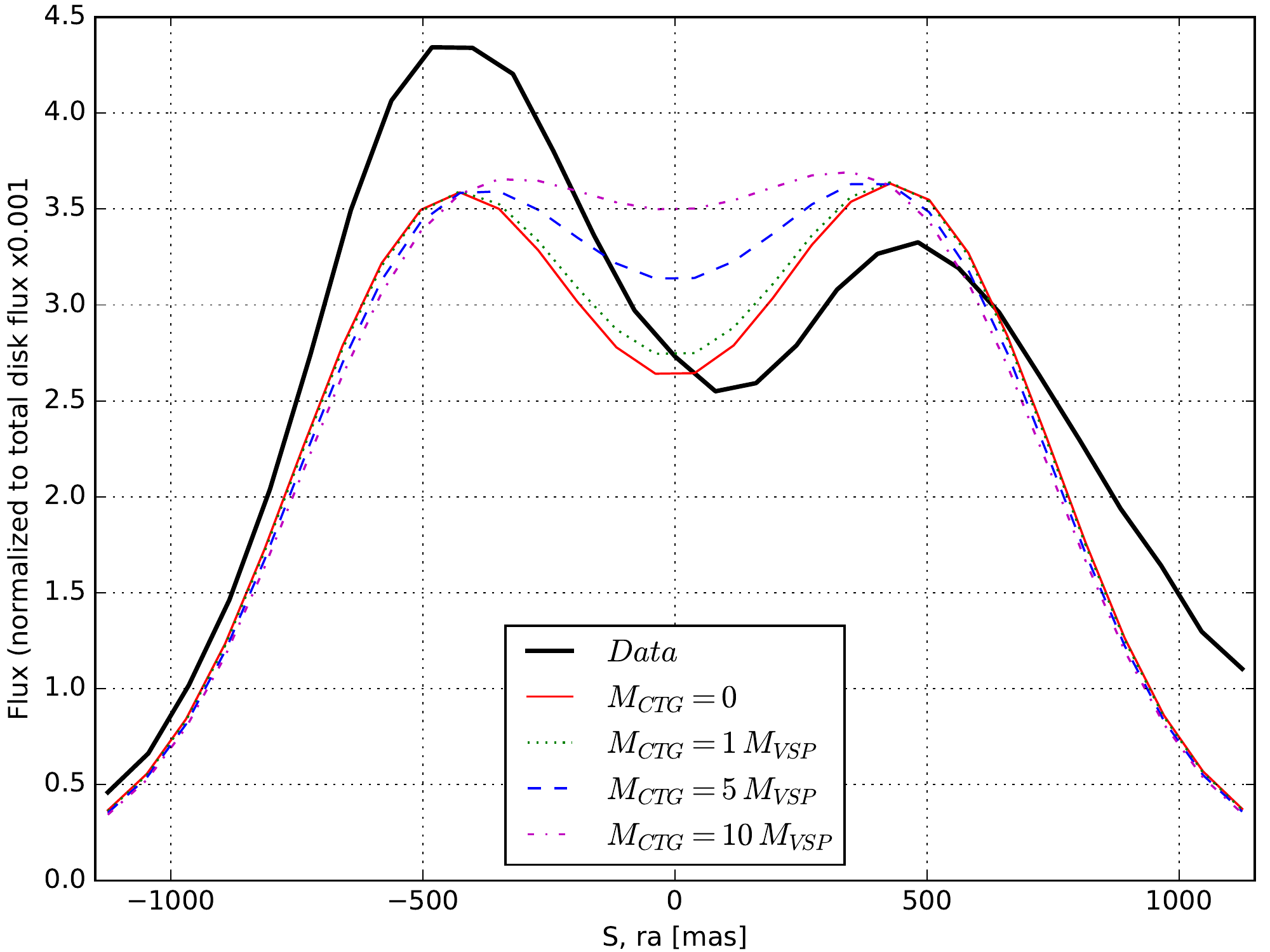}
    \caption{Cuts along the semi-major axis of the data-image (thick black line) at 18.7 $\mu$m and of the synthetic-images at he same wavelenght and computed for an increasing amount of Classical Thermal Grains (CTG) in the PAH ring (0, 1, 5 and 10 M$_{VSP}$ worth of CTG). Each flux-profile is normalized to the total flux of the disk in each image.}
  \label{figure:SIL_PAH}
\end{figure}

Figure~\ref{figure:SIL_PAH} compares intensity profiles along the semi-major axis of the observed and modelled disks at 18.7 $\mu$m. Several models were computed, where the amount of thermal grains (silicates (70\%) \& carbonaceous (30\%)) with sizes between 0.3 and 300 $\mu$m (mass power index of -3.5), are introduced in the VSP-ring as settled material (one fifth of the VSP-ring scale-height). Although our axi-symmetrical model cannot obviously reproduce the eastern asymmetry, it shows a distinctive dip at the star location corresponding closely to the 18.7 $\mu$m observed image. When the amount of classical thermal-grains increases, the inner-ring becomes brighter due to the efficiency of silicates grains to radiate at 18-19 $\mu$m. The central dip progressively disappears until the flux-profile is nearly flat, for a thermal-grains mass of \ap10 M$_{VSP}$.

Note that the VIS2 data fitting was not affected by the adding of Classical Thermal Grains (CTG), which emit a negligible amount of flux in Ks-, Lp- and Mp-band. Indeed, the emission of such CTG is mostly visible through the emission ``bumps'' of the silicate grains, at \ap9.5 and 18 $\mu$m where the SED is dominated by the outer-disk. Until an addition of 5 M$_{VSP}$ worth of CTG in the VSP-ring, the fit to the SED do not show any variation. From 10 M$_{VSP}$ however, the model starts to clearly over-estimate the SED fluxes between \ap8 and 15 $\mu$m. Refer to Section~\ref{section:PAH} for a discussion.

\section{Closure Phases Morphology}\label{section:t3morph}

T3 are especially robust to atmospheric turbulence and optical aberrations. They are however in essence only sensitive to asymmetries: a point-symmetrical structure would lead to null T3 signal. As a consequence, an equal-brightness binary star would not produce any T3 signal either.

In the data, T3 at all epochs vary between -12 and +9$^{\circ}$ in Lp-band, -18 and +15$^{\circ}$ in Mp-band, and slightly less in Ks-band. This highlights that strong asymmetries exist in the IRS-48 disk, at all wavelengths and epochs. T3 calculated on the radiative transfer model-image in all bands merely reach 0.5$^{\circ}$; a disk alone does not explain the T3 data.

\subsection{Modelling}

We fit N point-sources to the T3 data in order to highlight the main locations of the asymmetries in the disk. Such simplistic models allow analytical, hence fast, fitting to the data.

We make use of six analytical models: from one to three point-sources in addition to the central star, either with or without the disk modeled with MCFOST. The central star is set as a reference to a flux of 1 and in the middle of the field. Each point source is described with three parameters: 1) angular separation to the central-star within [20, 300] mas, 2) position angle (East of North) within [0, 360]$^{\circ}$, 3) relative flux within [$\frac{1}{200}$, 1] stellar flux.

The disk is scaled and has its flux normalized with respect to that of the central star. It is treated as a ``background image'' on which T3 are measured at the (U,V) coordinates of the data. Complex visibilities from each unitary model-object are added together and T3 values for the global model are obtained from the phases of the complex visibilities at the (U,V) coordinates of each T3 triangle.

The complex visibility $\mathcal{V}$ at the (U,V) coordinates for N point-sources is expressed by:
  \begin{equation}\label{eqn:viscomp}
     \Big(1 + \sum_{i=1}^N f_i\Big)~\mathcal{V} = 1+\sum_{i=1}^N f_i~\exp\Big(-\textbf{j}\frac{2\pi}{\lambda}~(U~\delta_i~+~V~\alpha_i)\Big),
  \end{equation}
where $\lambda$ is the central wavelength of the band. $\delta_i$, $\alpha_i$ are the angular separation and position angle of the $i^{th}$ point-source with respect to the central star (projected along declination and right ascension axes), and $f_i$ is its the flux ratio.

The error estimation of interferometric data is extremely challenging, due to the intrinsic nature of the measurements that go through heavy data-reduction. Systematic errors are calibrated using several PSF-references (thanks to the observation pattern ``Calibrator--Science Target--Calibrator``), while random errors are estimated from the scatter on the data. However, possibly large systematic errors remain even after calibration. The reason for this is a variation of experimental conditions between the measurements on the PSF-reference(s) and the science target: either turbulence (i.e. AO correction), airmass, position on the detector, position angle (i.e. angle-dependent aberrations in the optics).

\subsection{Fitting}

A thorough treatment of the variance-covariance matrix of T3 requires at least as many independent measurements of each target as independent T3 \citep{Ireland13}, which is in practice not feasible. Instead, the variance-covariance matrix can be modeled \citep{Kraus08}, or an additional error-term can be added in quadrature (i.e. sum of independent variances) to the error estimation on the data and tuned to normalize the reduced chi-square distribution of the fitting \citep{Hinkley11}. However, the non-linear form of the complex visibility expression to evaluate, see Eq.~(\ref{eqn:viscomp}), makes it impossible to determine the number of degrees of freedom; the reduced chi-square can not be used to compare models \citep{Andrae10}. Rather than tuning it using reduced chi-square, we choose to fit the additional variance-term $\sigma_{systematic}^2$ according to a maximum likelihood estimation. We define the total variance $\sigma_{i,total}^2$ to be:
  \begin{equation}\label{eqn:add_variance}
    \sigma_{i,total}^2 = \sigma_{i,random}^2 + \sigma_{systematic}^2
  \end{equation}
in order to account for errors-correlation and uncalibrated systematic errors of the T3.

We use a Markov-Chain Monte-Carlo (MCMC) simulation using the \textit{emcee} pure-Python implementation of Goodman \& Weare's Affine Invariant Ensemble sampler~\citet{ForemanMackey13} to maximize the full log-likelihood expression $ln\mathcal{L}$ defined by:
  \begin{equation}\label{eqn:loglike}
     -2~ln\mathcal{L} = \sum_i \Bigg[\frac{\big(Data_i - Model_i\big)^2}{\sigma_{i,total}^2} - \ln\Big(\frac{2\pi}{\sigma_{i,total}^2}\Big)\Bigg],
  \end{equation}
where the index $i$ iterates over each (U,V) coordinate of a dataset. While the first term under the sum favors high values of $\sigma_{i,total}$ (hence higher values of the fitted additional variance $\sigma_{systematic}$), the second term penalizes the increase of $\sigma_{systematic}$. The balance between these terms ensures the convergence of the estimation.

In order to compare models, we calculate the Bayesian Information Criteria (BIC) at the maximum likelihood of each model. The BIC is defined by~\citet{schwarz1978} such as:
  \begin{equation}\label{eqn:aiccrit}
     BIC_{Model} = 2~(k\ln(n) - ln\mathcal{L}_{Model}\mid_{max}),
  \end{equation}
where k is the number of free parameter in the model and n the sample size. The presence of k weighted with the sample size ensures that the more simple models will be preferred (to a certain extent) over more complex models. We estimate the relative probability of all models with respect to the null hypothesis of a unique central star without any disk using Likelihood Ratio Tests (LR-Test) in its most general formalism:
\begin{eqnarray}\label{eqn:likerattest}
      LR_{Model/null} &= \exp\Bigg(\frac{BIC_{null}-BIC_{Model}}{2}\Bigg)\nonumber\\
                      &= \frac{1}{n^{3N}}\frac{\mathcal{L}_{Model}\mid_{max}}{\mathcal{L}_{null}},
\end{eqnarray}
where N is the number of point-sources of the model.

\subsection{Results}

We compute the relative probability for all models on each of the six datasets, which we tabulate as log of Eq.~(\ref{eqn:likerattest}) in Table~\ref{table:model_compare}. All six datasets except Lp-band epoch 3 favor the 2 point-source model rather than the 0, 1 or 3 point-source models, and all models favor the presence of a disk over a disk-less star (with the exception of Ks-band epoch 1 where the relative probabilities of a star with and without a disk are nearly identical). Since the fits on all T3 datasets were performed independently, there is a very clear statistical significance of a model that comprises a disk and two additional point-sources. Note that given the relatively low flux from the point-sources compared to that of the central star and the disk altogether, the additional two point-sources do not impact significantly the fit of the radiative transfer disk model on the VIS2 data.

\begin{table}
  \caption{Logarithm of the relative probability for each model and all datasets with respect to the null hypothesis of a single star without disk. The highest value the better. A negative value tells that the model performs worse than the null hypothesis. Bold values highlight the most probable model for all datasets. The relative probability values for different datasets cannot be compared between each-other. n is the number of independent data point (15 per acquisition sequence, see Table~\ref{table:observation_nacosam}).}
  \begin{center}
  \begin{tabular}{ l c c c c}
   \hline\hline
   Model & NULL & 1PS & 2PS & 3PS \\
  
   \hline \multicolumn{5}{c}{Ks epoch 1, n=60} \\
   S & -- & -1.16 & \textbf{4.68} & -0.70 \\
   S+D & 0.01 & -1.36 & \textbf{4.64} & -0.02 \\
   \hline \multicolumn{5}{c}{Lp epoch 1, n=60} \\
   S & -- & 84 & 178 & 166 \\
   S+D & -0.46 & 96 & \textbf{223} & 223 \\
   \hline \multicolumn{5}{c}{Mp epoch 1, n=30} \\
   S & -- & 12.4 & 12.2 & 4.6 \\
   S+D & -2.8 & 15.1 & \textbf{15.3} & 10.2 \\
   \hline \multicolumn{5}{c}{Lp epoch 2, n=30} \\
  S & -- & 32 & 72 & 65 \\
   S+D & -0.25 & 38 & \textbf{108} & 87 \\
   \hline \multicolumn{5}{c}{Lp epoch 3, n=45} \\
   S & -- & 52 & 113 & 131 \\
   S+D & -0.21 & 60 & 146 & \textbf{185} \\
   \hline \multicolumn{5}{c}{Lp epoch 4, n=60} \\
   S & -- & 51 & 176 & 227 \\
   S+D & 0.06 & 59 & \textbf{215} & 210 \\
   \hline
   \end{tabular}\par
\end{center}
NULL$\equiv$ Null hypothesis i.e. single star without disk\\D$\equiv$Disk model-image in the corresponding band\\PS$\equiv$Point source model-objet
   \label{table:model_compare}
 \end{table}

Finding the most probable model among several does not however ensure that the fit, hence the model, is satisfactory. We define \textit{residuals} to be the normalized error of the model such that:
  \begin{equation}\label{eqn:residuals}
     Residuals_i = \frac{Data_i - Model_i}{\sigma_{i,total}},
  \end{equation}
where the index $i$ iterates over each (U,V) coordinate of a dataset. Table~\ref{table:t3_res} shows that the additional systematic error $\sigma_{systematic}$ in Ks- and Lp-band is \ap0.8$^{\circ}$, and reaches 4$^{\circ}$ in Mp-band. These uncalibrated systematic errors are consistent with previous performance and slightly worse SPHERE-SAM mode: \ap0.5$^{\circ}$ (SPHERE SAM commissioning team, private conversation). This table also shows that on all datasets, the RMS of residuals calculated using $\sigma_{i,total}$ is very close to unity, meaning this model does not under- or over-fit the T3 data. Note that the fact that RMS values of residuals is unity is a pure consequence of the likelihood maximization of the additional variance parameter, not an a-posteriori normalization. The average of residuals is only a fraction of unity, which denotes only a slight deviation from normal residuals (i.e. unbias).

\begin{table*}
  \caption{\textbf{Top Table}: Overview of the quality of the T3 data fit with the favored model (2 point-sources and disk), for all datasets. The total resulting error ($\langle\sigma_{i,total}\rangle$) defined by Eq.~(\ref{eqn:add_variance}), composed of the maximum likelihood estimation of the additional error term ($\sigma_{systematic}$), and the average random errors of the data ($\langle\sigma_{i,random}\rangle$). The RMS of residuals remains close to unity while its average close to zero, meaning no or little bias.
  \textbf{Bottom Table}: Maximum likelihood estimation of the point-sources parameters for the most probable model, consisting of two point-sources in addition to a central star with the disk found in the radiative transfer VIS2 data fitting, for all datasets.}
  \centering
  \begin{tabular}{ l | c c c | c c }
  \hline\hline
    Data set & $\langle\sigma_{i,random}\rangle$ ($^{\circ}$) & $\sigma_{systematic}$ ($^{\circ}$) & $\langle\sigma_{i,total}\rangle$ ($^{\circ}$) & $RMS(residuals_i)$ & $\langle residuals_i\rangle$ \\ \hline
    Ks epoch 1 & $1.6$ & $0.86\pm0.17$ & $1.82$ & $0.97$ & $0.35$ \\
    Lp epoch 1 & $0.27$ & $0.68\pm0.17$ & $0.73$ & $1.01$ & $0.20$ \\
    Mp epoch 1 & $1.9$ & $4.0\pm0.6$ & $4.5$ & $1.00$ & $0.10$ \\ \hline
    Lp epoch 2 & $0.44$ & $0.63\pm0.23$ & $0.77$ & $0.97$ & $-0.21$ \\ \hline
    Lp epoch 3 & $0.25$ & $0.86\pm0.11$ & $0.89$ & $1.00$ & $0.18$ \\ \hline
    Lp epoch 4 & $0.26$ & $0.97\pm0.11$ & $1.0$ & $1.00$ & $0.26$ \\
  \hline
  \end{tabular}
  \\
  \vspace*{0.5cm}
 
  \begin{tabular}{ l | c c c | c c c}

  \hline\hline
   & \multicolumn{3}{c}{First Point-Source} & \multicolumn{3}{|c}{Second Point-Source} \\
  Data set & Sep. (mas) & PA ($^{\circ}$) & $\Delta$mag & Sep. (mas) & PA ($^{\circ}$) & $\Delta$mag \\ \hline
  Ks epoch 1 & $101\pm4$ & $281\pm4$ & $4.9\pm0.3$ & $69\pm5$ & $16\pm8$ & $4.8\pm0.3$ \\
  Lp epoch 1 & $105\pm2.5$ & $270.3\pm1.1$ & $3.35\pm0.12$ & $41_{-3}^{+9}$ & $41\pm4$ & $2.7\pm0.5$ \\
  Mp epoch 1 & $89\pm11$ & $269\pm4$ & $2.9\pm0.3$ & $39\pm9$ & $52\pm12$ & $1.8_{-0.6}^{+1.0}$ \\ \hline
  Lp epoch 2 & $111\pm2.5$ & $275.1\pm1.3$ & $3.20\pm0.20$ & $34_{-2}^{+15}$ & $47\pm4$ & $1.8_{-1.0}^{+1.7}$ \\ \hline
  Lp epoch 3 & $109\pm3$ & $276.5\pm1.3$ & $3.40\pm0.05$ & $39_{-2}^{+10}$ & $44\pm4$ & $2.3_{-0.5}^{+0.7}$ \\ \hline
  Lp epoch 4 & $119\pm4$ & $284.6\pm1.1$ & $3.25\pm0.06$ & $50\pm4$ & $43.3\pm1.1$ & $2.72\pm0.22$ \\
  \hline
  \end{tabular}
  \label{table:t3_res}
\end{table*}

The results for the maximum likelihood parameter estimation for the favored model with a disk and two point-source are tabulated in Table~\ref{table:t3_res}. Although point sources were free to take any positions and fluxes, the parameter estimation locates both point sources at a consistent spot at all epochs and wavelength. The first point-source is \ap105 mas on the west of the star with a contrast ratio of \ap3.3 mag in L-band; the second point-source location is somewhat noisier due to its proximity to the star although still consistently found at \ap45 mas on the north-east with a contrast ratio of \ap2.5 mag in L-band. These locations correspond closely to the inner-rim of the VSP-disk.

\subsection{Point-sources Characterization}

Fitting point-sources to T3 data is equivalent to finding the photo-center of asymmetrically emitting regions; it does not constrain their size nor shape. Additional fits were carried using uniform-disks (three point-source parameters and an additional diameter), Gaussian blobs (three point-source parameters and a semi-major axis) and two-dimensional Gaussian blobs (three point-source parameters, two semi-major axes and a position angle) instead of point-sources. However, the data was not able to constrain well the additional parameters: the point-sources are mostly unresolved (i.e. angular extension $\lesssim$3 AU).

We determine the color-temperature of both point-sources at epoch 1, using their contrast ratios in Ks-, Lp- and Mp-bands. We assume a black-body behavior constrained by its effective temperature and its radiating surface, assumed circular and defined by its radius. We use \textit{emcee} to maximize the likelihood of the temperature parameter T taken in [100, 10000] K and the logarithm of the radius parameter R taken in [0.1, 40] R$_{\odot}$.

\begin{figure}[!]
  \centering
  \includegraphics[width=\hsize]{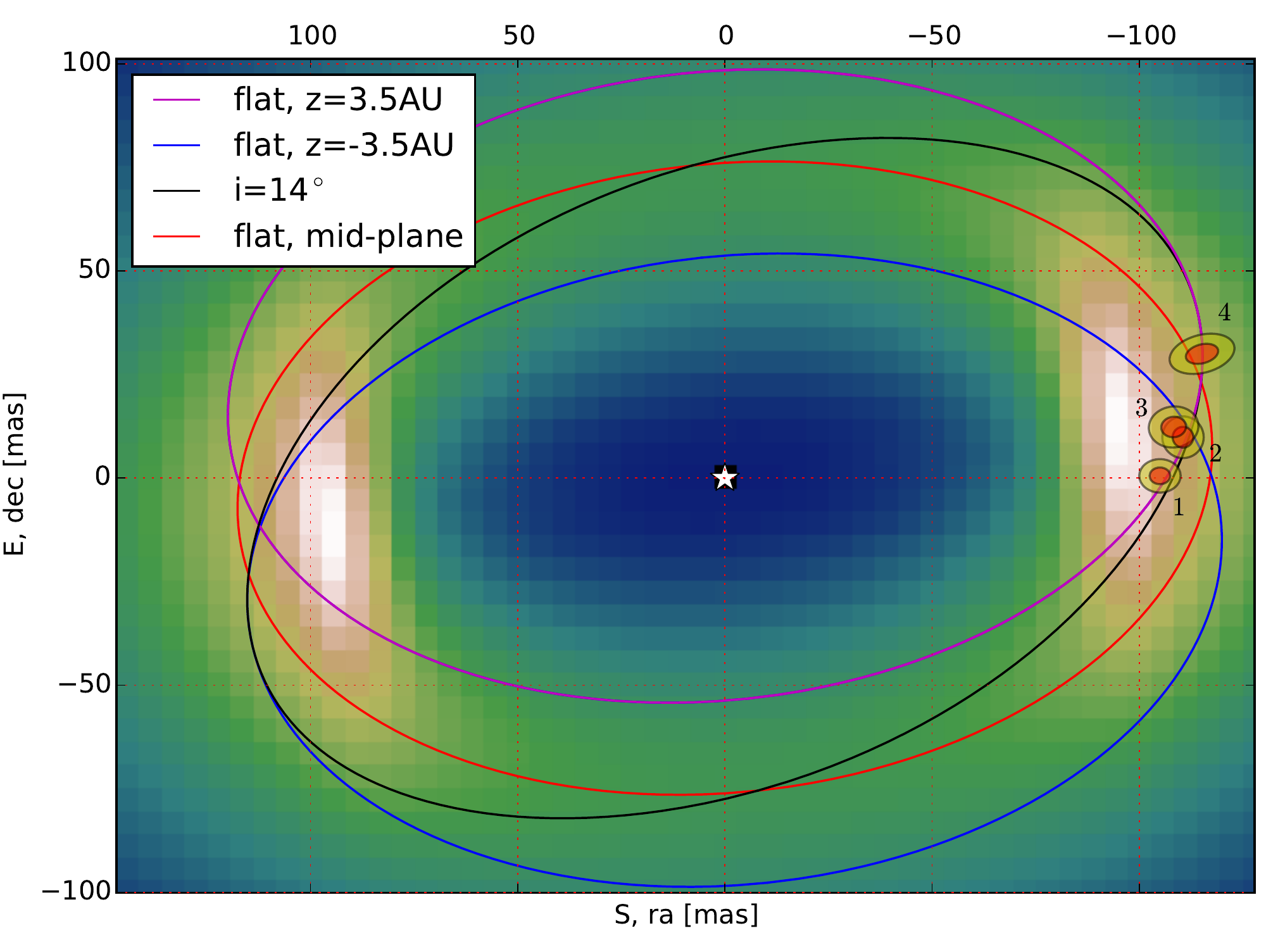}
  \caption{Locations of the first point-source in Lp-band through the four epochs (noted 1 to 4). Red and yellow ellipsoids are 1- and 2-sigma uncertainties for these locations. The background image represents the Lp-band model-image of the radiative transfer disk. It is shown in linear scale, the star was subtracted. Purple, red and blue lines represent a circular orbit of semi-major axis $a$=14.2 AU for a body located, respectively, 3.5 AU above the mid-plane, in the mid-plane and 3.5 AU below the mid-plane. The black line represents a 14$^{\circ}$ inclined circular orbit with same $a$.}
  \label{figure:companion}
\end{figure}

\begin{figure*}[!]
  \centering
  \includegraphics[width=0.5\hsize]{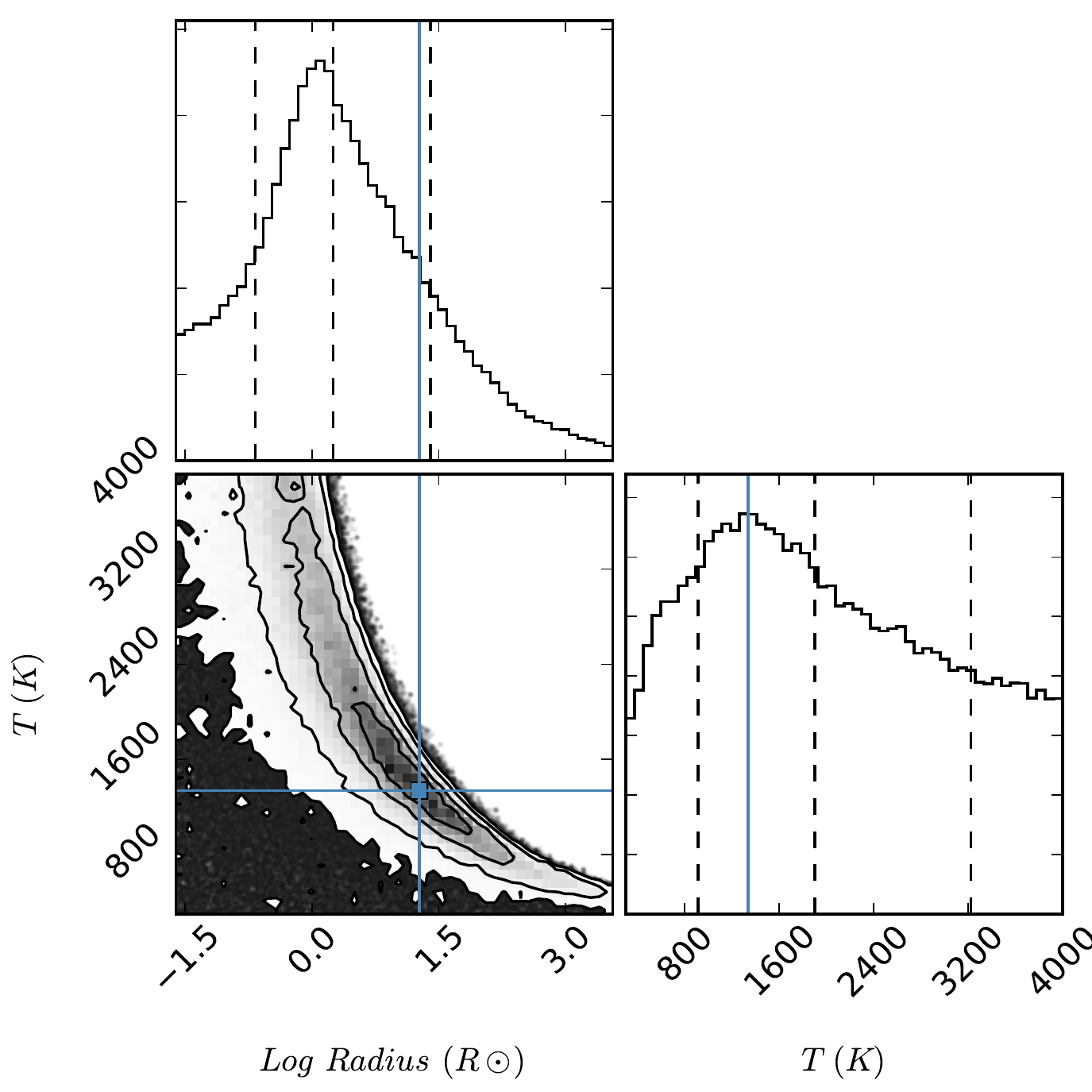}%
  \includegraphics[width=0.5\hsize]{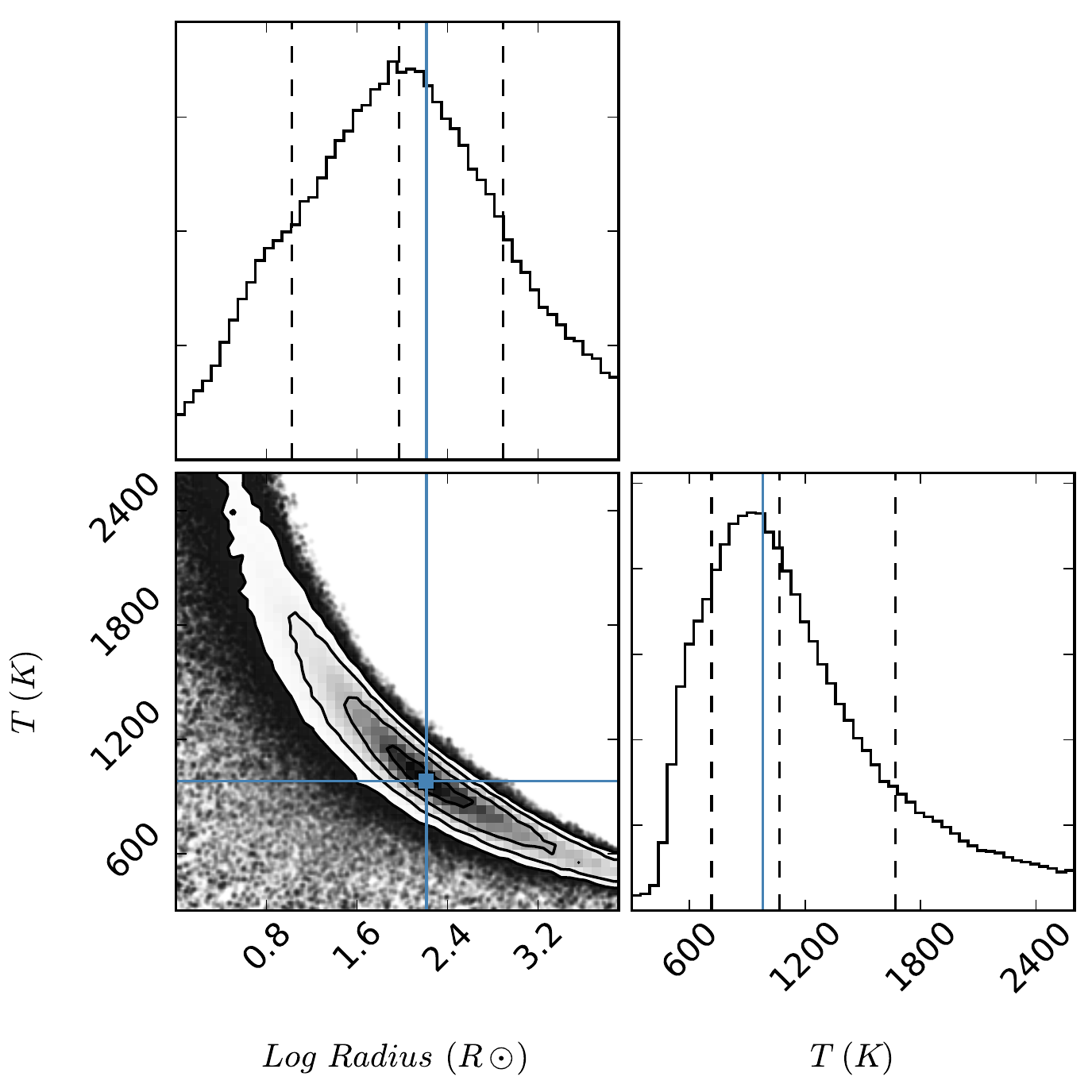}
  \caption{One- and two-dimensional histograms showing the variance, covariance and maximum likelihood (blue lines) of R and T parameters obtained from the fit. First point-source is on left, second one is on right. Dashed lines on one-dimensional histograms represent the 16$^{th}$, 50$^{th}$ (i.e. median) and 84$^{th}$ percentiles while iso-contours on two-dimensional histograms represent 0.5, 1, 1.5 and 2-sigma areas. The y-axes scaling on one-dimensional histogram is arbitrary; it shows the relative probability distribution function.}
  \label{figure:colortemp}
\end{figure*}

The maximum-likelihood color-temperature found for the first point-source is \ap1350 K, see Figure~\ref{figure:colortemp}. However, one-sigma error-bars yield several hundreds of Kelvin because of large error-bars on the point-source contrast ratios obtained with T3 data. Still, this temperature corresponds closely to PAH-temperatures found with radiative transfer modelling at the unprojected point source location, \ap12.5 AU; refer to Figure~\ref{figure:temperatures}.

The second point-source shows a somewhat lower temperature, most likely of \ap1000 K. Here again, one-sigma errors-bars yield several hundreds Kelvin large: from 600 to 1300 K, refer to Figure~\ref{figure:colortemp}. Due to its proximity to the star, at or below 0.5$\lambda/D$, the angular separation of the second point-source is particularly degenerate with its contrast ratio. However, the position angle remains mostly unaffected and shows no significant angular movement with epochs.

Discrepancy on the epoch 1 locations of the second point-source between the different bands may be due to several factors, given that fitting T3 data only gives the location of the photo-center of an asymmetrically bright region. First, this location may change with wavelength due to the intrisic cause of the asymmetry: a clump of segregated grains, an arm of material showing decreasing temperatures as it spirals out from the star, or visual effects of optical depth, opacity and shadows. Second, the data -- especially the sample size and the error-bars -- allowed to fit only two point-sources. This second point source may however well be the (poorly defined) photo-center of a more complex system of asymmetries, unresolved by the data.

The first point-source in Lp-band shows a clear rotational movement through different epochs. It is unlikely related to the (U,V) field rotation, since it shows much smaller variations over the four epochs, and not monotonic. This movement is compatible with a circular orbit in the VSP-disk and in the same direction as the gas kinematic found by~\citet{Brown12a,Bruderer14} (with South as the near side). If we assume a circular orbit, the origin of the over-brightness is located at a radius of \ap14.2 AU, and a height of \ap3.5 AU (see below). This over-brightness location corresponds to the densest region of the VSP-ring in our radiative transfer modelling, as Figure~\ref{figure:PAH_density_dust} tells. However, the nature of the over-brightness is not clearly constrained. It may be 1) an orbiting over-density of VSP material, or 2) the central star orbiting a center of mass, shifted from the center of the disk due to a massive companion.

Figure~\ref{figure:companion} shows the Lp-band point-source location in the disk as well as two circular-orbit solutions.
The first solution is an orbit with a semi-major axis of $a$=14.2 AU. It lies flat 3.5 AU above the mid-plane (hence inclination \ap0$^{\circ}$). This suggests that either the clump of material producing the over-brightness is mostly pressure-supported, or that only the top layer of it is detected. This latter option however implies that the center of mass and the photo-center of the clump are not co-located, inconsistent with the high transparency of this VSP-ring (see Figure~\ref{figure:tau_midplane}).
The second solution is an inclined orbit, $i$=14$^{\circ}$, with a similar semi-major axis, $a$=14.6 AU; this implies a gravity-dominated clump of material, orbiting the star.

In both orbital solutions, the over-brightness moves at an un-projected Keplerian velocity of 12.7$\pm1.9$ km.s$^{-1}$ in average over the four epochs (similar values are obtained for other intervals, between epochs 1-3 and 3-4); the total orbital period at this pace is \ap34 years. The theoretical Keplerian velocity at such distance and given the stellar mass is $\nu=\sqrt{G\times m_*/r}$=12.5 km.s$^{-1}$.

\section{Discussion}

\subsection{Age of IRS-48}\label{section:age}

Finding a brighter star, hence younger (4 Myr), partly solves an evolutionary puzzle on this target.

First, the discrepancy between the molecular Ophiuchus cloud age of \ap1 Myr \citep{Luhman99} and the stellar age is widely decreased, even though a fair difference remains that cannot be explained to date, as it is the case for all class II disks in Ophiuchus. IRS-48 might have been part of an earlier star-formation episode in Ophiuchus, as postulated by~\citet{Brown12a}.

Second, the disk of our model is entirely dust-depleted up to 55 AU, except for a thin VSP-disk between 11 and 26 AU. This implies that the evolutionary stage of the disk is probably more advanced than what was previously thought. Accretion was estimated at \ap10$^{-8.5}$ M$_{\odot}$/yr by \citep{salyk13,Follette15} using $Pa\beta$ and $Br\gamma$ lines. This non-negligible accretion rate raises the question of the origin of such material, and does not seem consistent with either the \textit{old} model (where one finds 8.0\pow{-12} M$_{\odot}$ dust in the inner-most disk, and 8.0\pow{-10} M$_{\odot}$ PAH until 50 AU), or our new model (3.7\pow{-10} M$_{\odot}$ VSP at 11-26 AU). Moreover, the disk was found to be gas-depleted until 20 AU, which puts a very hard constrain on how much material is to be found in the first tens of AU, available for accretion. This means that some active replenishment processes take place to sustain accretion, or that IRS-48 is on the verge of becoming a debris disk.
median disk lifetimes was estimated to be between 2 and 3 Myr, with an exponential decay of 2.5 Myr characteristic-time \citep{Mamajek09}. Studies show that an inner disk is detected through NIR excess in 60 to 80\% of stars younger than 1 Myr while no more than 10\% of stars older than 10 Myr seem to possess one \citep{Strom89}.

Finally, a younger age for IRS-48 together with a later disk-evolution stage is consistent with the fact that massive stars erode their disk at a larger pace than solar-type stars. Indeed, earlier studies on Upper Scorpius OB association \citep{Carpenter06,Williams11} showed that \ap20\% of 127 K- and M-stars are surrounded by optically thick disks while none of 30 F- and G-stars showed such disk for wavelength \ap16 $\mu$m, implying that primordial circumstellar disks are consumed faster by a massive star. This result is confirmed by~\citet{Ribas15} who show that only 2\% of M$\geqslant$2 M$_{\odot}$ stars aged $\geqslant$3 Myr show a protoplanetary disk.


\subsection{Disk Morphology, Stellar Binarity and Planetary Formation}\par

\subsubsection{30 AU CO gas-ring}

\citet{Brown12a,Bruderer14,vanderMarel16} find a \ap30 AU CO-ring with small radial extent, compatible with a Keplerian velocity for a 2.0 M$_{\odot}$ star with an inclination of 50$^{\circ}$. In addition, they find a gas-depleted region inside \ap20-25 AU.

Although our data is blind to gas -- consequently our model does not incorporate any gas structure -- the new IRS-48 model and the later disk-evolution stage postulated remain qualitatively compatible with these previous findings. Indeed, a photo-evaporation-dominated disk would show a clear depletion of gas and dust in the inner tens of AU. The VSP-ring from 11 to 26 AU acts as a very efficient shield of UV-radiation, effectively protecting gas further out from aggressive photo-evaporation. A thin ring of gas at 30 AU -- and potentially as close as 26 AU -- appears to be a consistent consequence of such configuration.


\begin{figure*}[!]
  \centering
  \includegraphics[width=\hsize]{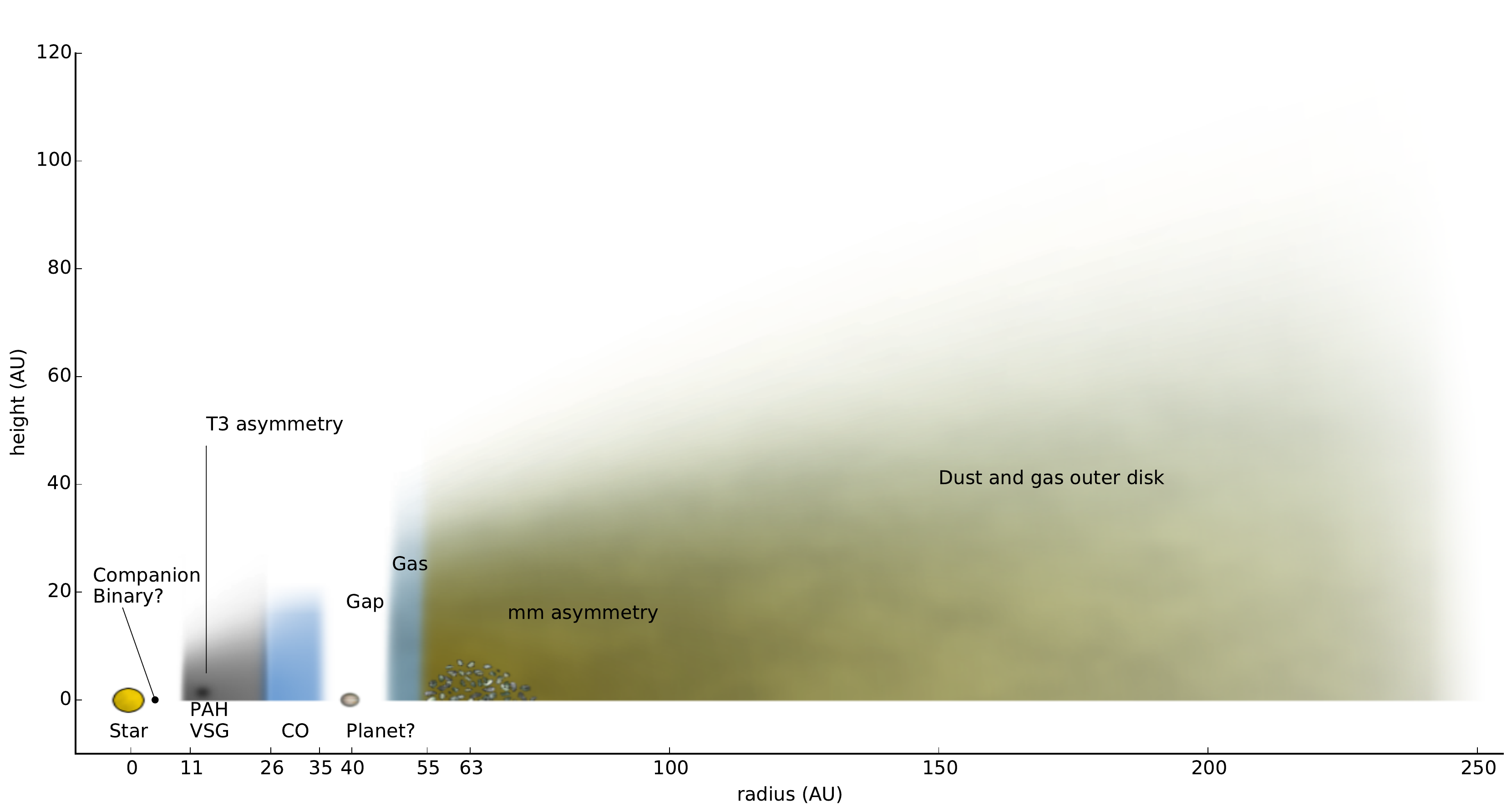}
  \caption{Schematic of the IRS-48 disk as presented in the new model. The outer-disk is from~\citet{Geers07}, the CO emission is from~\citet{Brown12a}, the mm-asymmetry from~\citet{vanderMarel13}, the inner 20 AU gas-depletion from~\citet{Bruderer14}. The proposed gas dip from 35 to 50 AU is proposed from~\citet{Bruderer14} findings on a first gas-depletion between 20 and 55 AU.}
  \label{figure:cartoon}
\end{figure*}

\subsubsection{VSP-ring Inner Truncation, Binarity}

\citet{Ghez93,Leinert93,Reipurth93} detect an over-abundance by a factor of two of binaries among the young stars in the Taurus region when compared to the results for the Main Sequence (MS) stars. Although further studies with larger populations slightly lower this over-abundance, it appears clear that YSOs have a binarity over-abundance compared to MS stars, to a degree that is however still under discussion as underlined by \citet{Ratzka05}.

Our data resolves a somewhat sharp truncation of the inner-rim at 11 AU. Besides photo-evaporation and accretion, a stellar companion with semi-major axis of order a few AU would reproduce the VSP-ring truncation at 11 AU \citep{Artymowicz94}. To the knowledge of the authors, no observation already performed in the visible or infrared, had the sensitivity to resolve such binary system. According to the PMS evolutionary models of~\citet{Siess00}, a K0-type star (5400 K) of 4 Myr has a luminosity of about 7.5 L$_{\odot}$ and a mass of \ap2.2 M$_{\odot}$. This luminosity remains small compared to that of IRS-48 (45 L$_{\odot}$): any stellar-type cooler than K0 would remain a nearly invisible companion for the visible and infrared instruments, especially at \ap25 mas typical spatial separations.

An indirect way to detect late-type companions is to check if such A-type stars emit in X\-ray. \citet{Stelzer06a,Stelzer06b} show that, given that A-type stars are not expected to have a corona and emit X\-ray, a high-energy emission from these stars is generally associated to late-type binarity. In the case of IRS-48, neither ROSAT and XMM archive data nor Chandra \citep{Imanishi01} showed X\-ray emission at this location. Although soft X\-ray would be absorbed by interstellar absorption, hard and extreme ones should be detected. Indeed, several very young stars (even Class I) are detected in X\-ray in the core-region of $\rho$ Oph by \citet{Ozawa05}. Given the completeness of several X\-ray studies as showed in their Figure 6, and the A$_v$=12.9mag for IRS-48, a 4 Myr old companion with T$_{eff}\gtrsim$3000 K (M6 star and brighter) should already have been detected around IRS-48.

Also, a heavy companion, $\gtrsim$1 M$_{\odot}$ (corresponding to a star brighter than \ap K6 stellar type on the 4 Myr isochrone) would reveal its presence through its mass, and the Keplerian velocities field it imposes on its circumbinary disk. \citet{Brown12a} find a 2 M$_{\odot}$ to explain a Keplerien rotation of a gas-ring at 30 AU ; this mass represents a lower-limit given that the gas-ring must have a sub-Keplerian velocity. IRS-48 does not likely have a bright and massive secondary star, and one can set a total binary mass at \ap4 M$_{\odot}$ and below.

IRS-48 is most probably not a binary star. However, a low-mass companion with M$\lesssim$0.5 M$_{\odot}$ and L$\lesssim$0.01 L$_{\odot}$ would be qualitatively compatible with X\-ray non-detection, CO 30 AU rotation speed, and with our model.

To date, the inner-truncatin at 11 AU is still to be explained.

\subsubsection{26-55 AU Cavity}

Both the presence of the outer-rim of the VSP-ring and the inner-rim of the outer-disk require the existence of an explanatory phenomenon. Based on the mm-asymmetry observed in the southern part of the disk, a 2 M$_{\odot}$ star and Hill radius estimates,~\citet{vanderMarel13} suggested the presence of a planet at 17-20 AU with M$\gtrsim$10 M$_{Jup}$.

If we carry similar first-order Hill radius estimates with the new model disk-structure and a higher-mass star, we speculate a M$_p$\ap3.5 M$_{Jup}$ planet at a radius of $a_p$\ap40 AU. Indeed,~\citet{DodsonRobinson11} find that the outer edge of the gas-cavity created by a planet is expected at \ap5 Hill radii $r_H$, defined by $3\times(r_H/a_p)^3 = M_p/M_*$. \citet{Pinilla12} find that for planets with mass between 1 and 3 M$_{Jup}$, dust dynamically accumulates at 7 $r_H$ from their location. With the later planetary parameters, we are able to obtain the dust-bump at 63 AU~\citep{vanderMarel13}, together with a VSP-ring bump at 17 AU, and gas densities peaking at 26 and 56 AU (all radii given as distances from the star). These radii correspond closely to the structure of our model; a schematic picture is shown on Figure~\ref{figure:cartoon}.

\subsection{PAH and VSG Evolution}\label{section:PAH}

\citet{Dullemond07} suggested that PAH and VSG do not settle well compared to classical thermal grains, and found that, under conditions of low turbulence, strong PAH emission-features can ``hide'' a substantial amount of settled thermal silicate grains.

We were indeed able to add up to five times the mass of the VSP-ring (5 M$_{VSP}$) worth of settled classical thermal grains the VSP-ring, without affecting the VIS2 data nor the SED fits.
However, adding such grains in the close vicinity of the star affects the MIR emission in such a way that the 18.7 $\mu$m model-image does not display the central dip with two bright spots at $\pm$55 AU from the star like the data-image does (Figure~\ref{figure:imgMIR}). Instead, a contiguously bright region lies along the whole semi-major axis of the disk.
This result shows that the VSP-ring most probably does not contain more thermal grains than 5$\times$M$_{VSP}$\ap2\pow{-9} M$_{\odot}$.

The presence of PAH in many circumstellar disks reveals that in these objects the dust coagulation process was apparently not effective enough to remove the PAH from these disks, or that some other process continuously replenishes PAH grains \citep{Dullemond07}. Out of theoretical and laboratory works,~\citet{Jochims94} showed that the largest and most regular PAH species are extremely stable to destructive radiations, i.e. photo-evaporation. Indeed, large PAH of 20-30 carbon atoms will preferably relax through PAH emission in bands rather than photo-fragmentation. As a comparison, one of the most stable PAH species, circumcoronene C54H18, has a typical size of 4.87 \AA. PAH used in our model span 4 to 10 \AA~(20 to 100 carbon atoms); VSG span 10 to 20 \AA~(100 to 1000 carbon atoms), see~\citet{Tielens08}.

\citet{Geers06} find a PAH-to-dust mass fraction of 6\% based on the abundance of 5\pow{-5} carbon atoms locked in PAH molecules per H nuclei (C$_{VSP}$/H). If we take updated values for PAH and VSG abundances from~\citet{Tielens08} (their Table 2), we find C$_{VSP}$/H=3.5\%. Our model showing a VSP-to-dust mass fraction $>$20\% in the VSP-ring between 11 and 26 AU clearly highlighs a depletion of a factor of \ap5-6 of classical dust grains up to 0.3mm compared to very small particles.

We are left with a problem where one shall study the disappearance rate of VSP through grain-growth, radiation pressure or inward drift, versus the rate of some replenishment process, e.g. the collisional destruction of larger grains or inward drift from an outer reservoir.

There is presently little experimental data on the size distribution resulting from sub-micrometer grain-grain collisions. \citet{Tielens08} report that theoretical studies on grain-grain collisions result in a mass distribution with power index of \ap-3.3. Although the minimum size is not constrained in this later study, given the layered structure of graphitic materials it is expected that the smallest fragments likely are small, two-dimensional structures (e.g. PAH-like molecular species). Further processing by atomic reactions, as well as UV processing, may then quickly transform these species into compact PAH \citep{Tielens08}. In the frame of interstellar medium shocks, \citet{Jones96} calculate that the catastrophic destruction (i.e. complete destruction of the grains, rather than cratering) of a 1000 \AA~grain by a 50 \AA~grain needs a critical velocity of approximately 75 km.s$^{-1}$; larger grains need larger collisional energy. They also conclude that interstellar grain-growth processes shall be 
much more efficient than catastrophic destruction in order to explain the observation that most of the interstellar dust comes in grains larger than \ap1000 \AA.

The orbiting velocity of the first point-source in the VSP-ring around IRS-48 is \ap12.7 km.s$^{-1}$, similar to the Keplerian velocity at that distance for the 2.5 M$_{\odot}$ star. Moreover, A0 stars do not generally show strong winds or emit high-energy radiations. It seems very unlikely that IRS-48 possesses a high catastrophic collision rate within its VSP-ring at 11-26 AU. It is therefore expected that PAH and VSG cannot be replenished easily through grain-grain collisions in the VSP-ring, in spite of a possible co-located reservoir of classical thermal grains up to five times the mass of VSP.

The radiation pressure estimate of the star can be achieved by calculating the stellar constant W\ap45,000 W.m$^{-2}$ at 1 AU. The radial acceleration from radiation pressure produced on a particle at radius r is obtained with a$_p$=W S$_{eff}$ c$^{-1}$.r$^{-2}$, with S$_{eff}$ the effective cross-section of the particle and c the speed of light in vacuum. It is found to be 39 (resp. 42, 17) times larger than the gravitational potential (G.M$_*$.r$^{-2}$) on the neutral PAH particles (resp. ionized PAH, VSG). This probably shows that the inward drift of such small particles is very low. The previous IRS-48 stellar model with radius 1.8 R$_{\odot}$ shows a somewhat lower but still overwhelmingly strong radiation pressure compared to gravitational potential for these species: 16, 17 and 7 times larger in average for neutral, ionized PAH and VSG.

The fact that the VSP-ring was observed to be consistently similar at a radius between \ap11-26 AU through the four epochs of observation spanning two years time, highlights that the VSP-ring is governed by a much more complex balance of physical processes. In order to remain observationally in-place, it must either be 1) constantly replenished by the outer-reservoir outside 55 AU -- possibly channeling triggered by the 40 AU companion -- or 2) kept in place as-is through 2a) gravitational shepherding from additional companions, or 2b) coating of large grains -- unaffected by radiation presure -- by PAH molecules.

\begin{acknowledgements}
We wish to thank Ewine van Dishoeck for her insight on SMA data, Vincent Geers for providing the VISIR NIR images, and Daniel Rouan for our talks on PAH and VSG. We also thank the Paranal staff for their support during the observations.
This work was supported by the French National Agency for Research (ANR-13-JS05-0005) and the European Research Council  (ERC-STG-639248).
Based on observations collected at the European Southern Observatory (ESO) during runs 086.C-0497(A), 087.C-0450(B), 088.C-0527(A), and 089.C-0721(A).
\end{acknowledgements}

\bibliography{irs.bib}

\end{document}